\documentclass[letterpaper,11pt]{article}
\pdfoutput=1
\pdfoutput=1
\pdfoutput=1

\usepackage{jheppub}
\usepackage{multirow}

\usepackage{subfig}
\usepackage{xspace}
\usepackage[countmax]{subfloat}

\usepackage{tikz}
\usetikzlibrary{math,calc,snakes, patterns,angles,quotes,arrows.meta,shapes.misc,decorations.pathmorphing,decorations.markings}

\usepackage{amssymb}
\usepackage{amsmath}
\usepackage{cancel}
\usepackage{tabu,booktabs}
\usepackage{color}
\usepackage{braket}
\usepackage{graphicx}
\graphicspath{ {./figures/} }
\usepackage{multirow}
\usepackage{verbatim}
\usepackage{amsthm}
\usepackage{slashed}
\usepackage{wasysym}
\usepackage{simplewick}
\usepackage{mathtools}
\usepackage{soul}
\usepackage{xspace}

\definecolor{verde}{RGB}{1, 113, 0}

\usepackage{enumitem}

\usepackage[only,llbracket,rrbracket]{stmaryrd}

\def\cC{\mathcal{C}}

\def\cN{\mathcal{N}}
\def\cO{\mathcal{O}}

\def\nn{{\nonumber}}

\newcommand{\namedref}[2]{#1~\hyperref[#2]{\ref*{#2}}}
\newcommand{\secref}[1]{\namedref{Section}{#1}}
\newcommand{\appref}[1]{\namedref{Appendix}{#1}}

\newcommand{\figref}[1]{\namedref{Figure}{#1}}

\def\be#1\ee{\begin{align}#1\end{align}}

\def\({\left(}
\def\){\right)}
\def\[{\left[}
\def\]{\right]}

\def\a{\alpha}

\allowdisplaybreaks[3]

\newcommand{\eps}{\epsilon}

\usepackage[font={small}]{caption}

\newcommand{\eeq}{\end{equation}}

\newcommand{\eeqq}{\end{equation*}}

\newcommand\eeqaa{\end{eqnarray*}}

\newcommand\eeqa{\end{array}}

\newcommand{\eea}{\end{eqnarray}}

\renewcommand{\Re}{\operatorname{Re}}
\renewcommand{\Im}{\operatorname{Im}}

\newcommand{\G}{\Gamma}

\newcommand\ea{\mathsf{a}}
\newcommand\eb{\mathsf{b}}

\def\njd{n_J^{(d)}}
\def\njdnID{\tilde{n}_J^{(d)}}
\def\Pjd{P^{(d)}_J}
\def\Mgap{M}

\def\Tpsiab{T_{\psi_{\ea,\eb}}}
\def\That{\widehat T}

\newcommand{\xavg}[1]{\llbracket #1 \rrbracket}

\usepackage[normalem]{ulem}

\usepackage{graphicx} 

\usepackage[section]{placeins}

\graphicspath{ {./figures/}}

\title{What is the graviton pole made of?}

\author[a,b,c]{Kelian H\"aring}
\author{and}
\author[c]{Alexander Zhiboedov}

\affiliation[a]{Institute for Theoretical Physics, University of Amsterdam, 1090 GL Amsterdam, The Netherlands}
\affiliation[b]{Fields and Strings Laboratory, Institute of Physics, École Polytechnique Fédéral de Lausanne (EPFL), CH-1015 Lausanne, Switzerland}
\affiliation[c]{Theoretical Physics Department,
	CERN, 1211 Geneva 23, Switzerland}

\abstract{We explore the physical mechanisms responsible for generating the graviton pole in twice-subtracted dispersion relations, both in flat space and in AdS. To characterize these mechanisms, we analyze the energy scale at which the graviton pole is generated in scattering experiments at various impact parameters. At large impact parameters, we identify the eikonal model
of high-energy gravitational scattering as a universal mechanism that generates the graviton pole in dispersion relations. At smaller impact parameters, the graviton pole can arise from stringy higher-spin resonances. The length scale at which the graviton pole generation scale departs from its semiclassical eikonal value indicates the breakdown of gravitational EFT. In flat space, we derive a Tauberian theorem for the graviton pole, which must be satisfied by any UV completion of gravity that admits twice-subtracted dispersion relations. In AdS, free or non-holographic CFTs offer an alternative mechanism to generate the graviton pole. More broadly, we find that the existing picture of high-energy gravitational scattering, including phenomena such as black hole formation and various stringy effects, is compatible with the twice-subtracted dispersion relations.
}

\begin{document}

\begin{flushleft}
	\hfill \parbox[c]{40mm}{CERN-TH-2024-178}
\end{flushleft}

\maketitle

\section{Introduction}

\begin{figure}[htbp]
    \centering
    \includegraphics[scale=0.9]{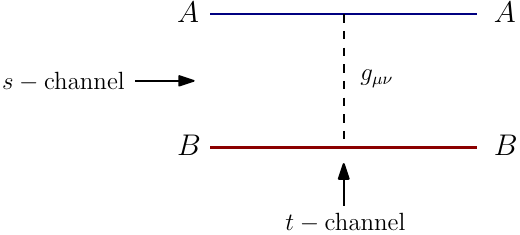}
    \caption{Universal gravitational attraction in elastic scattering due to the graviton exchange between particles $A$ and $B$.
    }
    \label{fig:treelevel}
\end{figure}

In the presence of gravity, elastic four-point amplitudes exhibit a graviton pole in the forward limit
\begin{equation}\label{eq:assymptotic_t0}
    \lim_{t \to 0} T(s,t) = \frac{8\pi G_N \gamma(s)}{-t} ,
\end{equation}
where the residue of this pole in general relativity takes the form \cite{DiVecchia:2023frv}
\begin{equation}
     \gamma(s) = \left(s-m_A^2-m_B^2\right)^2 - \frac{4m_A^2 m_B^2}{d-2}\, ,
\end{equation}
and $d$ is the number of spacetime dimensions. This universal graviton exchange graph is represented in \figref{fig:treelevel}. In the low-energy limit $s \to (m_A+m_B)^2$, the formula above captures Newton's law of gravitational attraction between bodies. For scattering at high energies $s \gg (m_A+m_B)^2$, we have $\gamma(s) \sim s^2$ since the exchanged graviton has spin two. For an interaction mediated by the spin $J$ particle, we would instead get $s^J$.

The case of a graviton, $J=2$, is special. On the one hand, only for $J \leq 2$ is it possible to construct simple, effective field theories of interacting massless particles in the IR \cite{Weinberg:1964ew}. On the other hand, the basic principles of unitarity and causality suggest that only the terms $s^J$ with $J \geq 2$ in the expansion of the amplitude are dispersive; namely, they can be represented by an integral over the discontinuity of the amplitude. This is a statement that relativistic amplitudes admit the twice-subtracted dispersion relations. In quantum field theory, this fact was established rigorously long ago \cite{Goldberger:1960rgk,Jin:1964zza}. For gravitational theories in flat space, the assumptions that go into the twice-subtracted dispersion relations have been recently discussed in \cite{Haring:2022cyf}. For gravitational theories in AdS, dispersion relations for the four-point function of local operators were studied in \cite{Caron-Huot:2017vep,Carmi:2019cub,Penedones:2019tng,Mazac:2019shk}. The flat space limit of AdS dispersion relations has been shown to lead to the twice-subtracted dispersion relations in \cite{Caron-Huot:2021rmr}. In this paper, we will assume that the twice-subtracted dispersion relations hold.

More precisely, in the presence of gravity, we can write the following equation
\be
\label{eq:dispersivepole}
\frac{8\pi G_N}{-t} + \cO(t^0,G_N^2) = 2\int_{-t}^\infty \frac{d s'}{\pi} {T_s(s',t) \over  (s')^3} ,~~~ t<0 \ , 
\ee
where $T_s(s,t) \equiv \lim_{\eps \to 0} {T(s+i \eps,t)-T(s-i \eps,t) \over 2 i}$ is the discontinuity of the elastic four-point amplitude, and for simplicity, we only wrote the leading low-energy contribution coming from the tree-level graviton pole.\footnote{We can focus on the full tree-level $O(G_N)$ scattering amplitude and not just the pole. This will change the equation, and we will do it in the bulk of the paper. In AdS, instead of the elastic amplitude, we consider the `elastic' four-point correlator $\langle \phi \phi \psi \psi  \rangle$.}
This equation, which can be derived starting from the twice-subtracted dispersion relations, states that the $t$-channel graviton exchange must be reproduced by exchanging degrees of freedom in the $s$-channel.\footnote{This statement is, of course, well-known from the tree-level scattering in string theory \cite{Veneziano:1968yb}. However, as we will explain below, the mechanism for the graviton pole generation in the full nonperturbative theory \emph{must} be different due to unitarity \cite{Soldate:1986mk}.} We would like to understand what are the possible physical mechanisms to generate the graviton pole in this equation, or, in other words, what is the graviton pole made of.

To explore the consequences of \eqref{eq:dispersivepole} it is convenient to smear it over the exchanged momenta
\begin{equation}\label{eq:SmearedDef}
    T_{\psi_{\ea}}(s) \equiv\int_{0}^{q_0} dq q \psi_{\ea}(q)\, T(s, t=-q^2)\, .
\end{equation}
Physically, this corresponds to performing scattering experiments with finite support in the impact parameter space $b \lesssim {1 \over q_0}$. The advantage of applying this functional is that by choosing an appropriate $\psi_a(q)$, we make the RHS of \eqref{eq:dispersivepole} \emph{nonnegative} thanks to unitarity \cite{Caron-Huot:2021rmr}.\footnote{For $t<0$, $T_s(s,t)$ is not necessarily nonnegative, and it is difficult to explore the consequences of \eqref{eq:dispersivepole} systematically because of the possibility of the cancellations in the dispersive integral.}  A simple family of functionals $\psi_{\ea}(q)$ that achieves this and allows us to explore the graviton pole takes the form
\be \label{eq:defPsiIntro}
\psi_{\ea}(q) =  \(\frac{q}{q_0}\)^\ea\(1-\frac{q}{q_0}\)^{(d-1)/2}, ~~~ 0<\ea\leq d-4 \ .
\ee
An important point is that upon acting with this functional, the ${8 \pi G_N \over -t}$ pole becomes the ${8 \pi G_N \over \ea}$ pole as we send $\ea \to 0$. However, the dispersive integral is now nonnegative, and therefore, a rigorous consequence of the sum rule can be derived using the standard Tauberian techniques \cite{korevaar_tauberian_2004}. The result is that to reproduce the graviton pole, the partial waves at large impact parameters must satisfy
\be
\label{eq:taubtheo}
\int_{0}^\infty {d s \over s^2} \Im a(s,b) \sim \frac{1}{2} \pi ^{3-\frac{d}{2}} \Gamma \left(\frac{d-4}{2}\right) {G_N \over b^{d-4}} \ , 
\ee
where $\Im a(s,b) \geq 0$ are the partial waves in the impact parameter space, and $\sim$ means equal on average (see \appref{sec:convention} for our conventions and \eqref{eq:tauberianImpactb} in the main text for the precise statement of the theorem). For a given impact parameter $b$, we can now define \emph{the graviton pole generation scale} $s_*(b)$ as an energy scale at which $\int_{M^2}^{s_*(b)} {d s' \over s'^2} \Im a(s,b)$ approximately satisfies \eqref{eq:taubtheo}.

\begin{figure}[htbp]
    \centering
\includegraphics[width=0.7\textwidth]{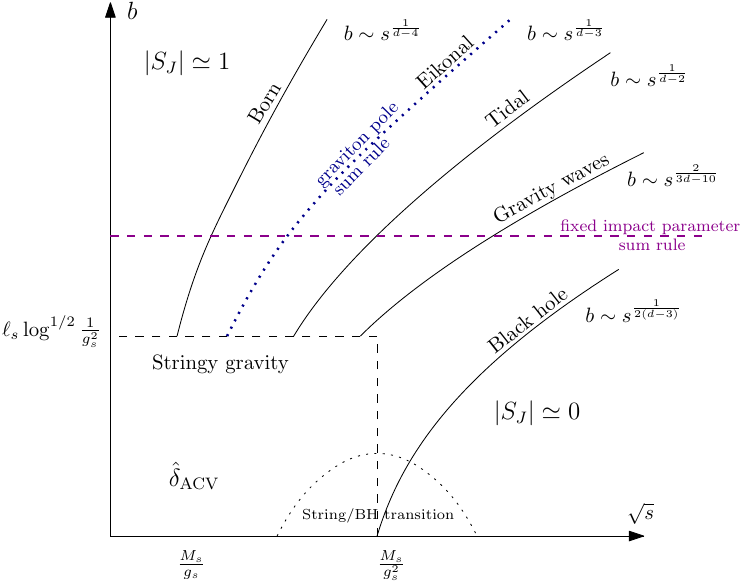}
    \caption{A schematic picture of the high-energy scattering in string theory according to ACV. At very large impact parameters, the scattering is essentially elastic; conversely, at small impact parameters, it becomes inelastic. Various effects scale with energy, as depicted. The stringy regime extends by an extra factor of $\log^{1/2} \frac{1}{g_s^2}$, which is the manifestation of the transverse spreading of the string. Eikonal scattering is responsible for generating the graviton pole in the dispersion relations, and it controls the nonperturbative high-energy behavior of the amplitude at fixed $t$. Thanks to the fact that other effects (e.g. tidal excitations, gravity waves emission, black hole formation) grow slower with energy, they decouple at high energy, and the leading high-energy asymptotic of the scattering amplitude is universal. In the weakly coupled regime $g_s \ll 1$, we can explore fixed impact parameter sum rules, for which the left-hand side of \eqref{eq:dispersivepole} is captured by the tree-level amplitude. In this case, the graviton exchange can be generated by stringy effects (or higher spin resonances) for $b \lesssim \ell_s \log^{1/2} {1 \over g_s^2}$. The dashed line about the string/black hole transition regime signifies that we do not have good control over scattering there.}
    \label{fig:ACV}
\end{figure}

To explore \eqref{eq:taubtheo} and various ways in which it can be satisfied, we use the general picture of high-energy gravitational scattering as developed in a series of papers by Amati, Ciafaloni, and Veneziano (ACV) \cite{Amati:1987uf,Amati:1987wq,Amati:1988tn,Amati:1990xe}, see also \cite{tHooft:1987vrq,Muzinich:1987in,Ademollo:1990sd,Giudice:1998ck,Giddings:2007bw,Giddings:2009gj}.
This picture was derived in the context of string theory by realizing that high-energy scattering in gravity admits a simple semiclassical description in the impact parameter space. We summarize the physical picture of ACV in \figref{fig:ACV}. By changing $b$ in the equation above, we can effectively perform scattering experiments at different impact parameters $b$ and, therefore, test the consistency between the ACV picture of high-energy gravitational scattering and the twice-subtracted dispersion relations.

\begin{figure}
    \centering
    \includegraphics[width=0.7\linewidth]{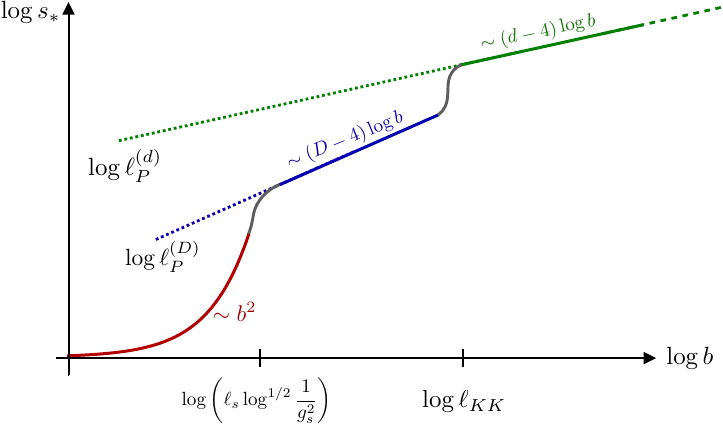}
    \caption{The graviton pole generation scale $s_*(b)$ as a function of the impact parameter $b$. We imagine a string theory with large extra dimensions and a low string scale. $\bullet\,$For large impact parameters $b$, the graviton is produced by the eikonal scattering with the characteristic energy scale being $\frac{G_N^{(d)} s_*^\text{eik}}{b^{d-4}} = O(1)$.
    $\bullet\,$As we decrease impact parameters below the scale of extra dimensions $b \lesssim \ell_{KK}$, a new scale emerges which is controlled by the $D$-dimensional Planck scale $\frac{G_N^{(D)} s_*}{b^{D-4}} = O(1)$. $\bullet\,$For $b \lesssim \ell_s \log^{1/2}\frac{1}{g_s^2}$, the string corrections become important, and the graviton is effectively generated by the tree-level string amplitude at energies ${\ell_s^2 \log \ell_s^2 s_* \over b^2} \sim O(1)$.  The length scale $\ell$ at which $s_*(\ell) \ll s_*^\text{eik}(\ell)$ signifies the breakdown of low-energy gravitational EFT, which controls the large impact parameter expansion of the $d$-dimensional eikonal operator. On the plot above, it corresponds to the Kaluza-Klein scale $\ell_{KK}$, at which point the relevant description becomes in terms of the $D$-dimensional eikonal scattering. On the other hand, the length scale $b_{\star}$, at which the effective spacetime dimensionality probed in the fixed impact parameter scattering experiment ${\cal D}(b) = {\partial \log s_*(b) \over \partial \log b}$ departs from its semiclassical value $D-4$ is the length scale at which gravitational EFT breaks down, i.e. the species scale \cite{Dvali:2007hz,Dvali:2007wp,Dvali:2008ec,vanBeest:2021lhn,Harlow:2022ich} or the higher-spin onset scale \cite{Caron-Huot:2024lbf}. In the plot above it corresponds to either the string scale, or the Planck scale.}
    \label{fig:gravScaleSch}
\end{figure}

Based on the analysis of gravitational scattering in string theory, we observe that there are two basic mechanisms to generate the graviton pole in \figref{fig:gravScaleSch}: via higher spin resonances (stringy modes) at small impact parameters or through eikonal scattering at large impact parameters. The simplest and universal mechanism for dispersively generating the graviton pole is provided by eikonal scattering. In this case, no new degrees of freedom are necessary, and the scattering amplitude is purely elastic at high energies. 
It is remarkable that Newton's potential, or the graviton pole that tells apples how to fall, is produced by inserting the Shapiro time delay into the twice-subtracted dispersion relations. The Shapiro time delay arises from propagation through the gravitational shockwave \cite{Aichelburg:1970dh,Dray:1984ha,Camanho:2014apa}, and governs the discontinuity of the amplitude $T_s(s,t)$ at large $s$.

As we will explicitly check, this mechanism works in flat space and AdS. 
The second mechanism is specific to string theory and is realized at small impact parameters, where the graviton is generated through higher spin resonances. Importantly, the tree-level string theory amplitude that achieves it violates nonperturbative unitarity; therefore, other physical effects should kick in to fix this problem. This is precisely what happens in the ACV picture around $\sqrt{s} \sim {M_s \over g_s}$, where inelastic stringy effects become important.
Finally, if we have extra dimensions with a characteristic scale $\ell_{KK} \gg \ell_{s}$, the graviton pole generation mechanism transitions to the higher-dimensional eikonal scattering, see \figref{fig:gravScaleSch}. As we will explicitly check, the same discussion readily applies to gravitational scattering in AdS and its CFT dual. It is interesting to ask whether \eqref{eq:taubtheo} can be satisfied by integrating out QFT degrees of freedom. It is impossible to generate the graviton pole within the framework of local QFT \cite{Weinberg:1980kq}.\footnote{Here we are talking about QFT degrees of freedom that populate the same spacetime as gravity. Of course, gravity can emerge from QFT degrees of freedom holographically \cite{tHooft:1993dmi,Susskind:1994vu,Maldacena:1997re,Gubser:1998bc,Witten:1998qj}.} For gapped QFT coupled to gravity, the answer is obviously no because, in this case, the partial waves quickly decay with spin and cannot satisfy \eqref{eq:taubtheo}, see, e.g., \cite{Arkani-Hamed:2020blm,Bern:2021ppb}. For massless free theories, it has also been recently shown to be impossible \cite{Caron-Huot:2024lbf}. 

The dispersion relations for the graviton pole have been extensively studied in the literature. The presence of the pole was an obstacle to deriving bounds on Wilson coefficients, where one usually expands the dispersion relation at small momentum transfer $t$. To circumvent this technical problem, the authors \cite{Caron-Huot:2021rmr,Caron-Huot:2022ugt, Caron-Huot:2022jli} introduced a smearing similar to \eqref{eq:SmearedDef} and recovered positivity. This allowed them to obtain bounds on the Wilson coefficients without introducing new assumptions. This method was later used in various systems containing gravitons  \cite{Henriksson:2022oeu,Hong:2023zgm, Albert:2024yap, Beadle:2024hqg, Xu:2024iao}. Another option was put forward in \cite{Bellazzini:2019xts}, where the authors compactified one spatial direction on a circle. This regulates the pole and allows for the derivation of bounds in the presence of the regulator. By studying the $\cO(G_N)$ gravitational amplitude, the authors \cite{Tokuda:2020mlf} introduced a Regge ansatz $T_s(s,t)\sim s^{2+ \alpha' t +\dots}$ for the amplitude which reproduces the graviton pole when inserted in the dispersion relation  \eqref{eq:dispersivepole}.
By explicitly subtracting it from the left-hand side of the dispersion relations, 
one can take the $t \to 0$ limit and proceed as usual for the positivity bounds. This idea was pursued in a series of papers and applied in different situations \cite{Loges:2020trf,Fuks:2020ujk,Herrero-Valea:2020wxz,Alberte:2020jsk,Noumi:2021uuv,Aoki:2021ckh,Herrero-Valea:2021dry,Hamada:2023cyt,Aoki:2023khq,Loc:2023nrp}. Later, this idea was reversed and the authors of \cite{Alberte:2021dnj,Herrero-Valea:2022lfd,deRham:2022gfe,Noumi:2022wwf,Caron-Huot:2024tsk} used dispersion relations to constrain UV parameters using information from the IR.  
In \cite{Caron-Huot:2024lbf}, the graviton pole sum rule was used to analyze the constraints on the low-energy matter content and identify the scale of breaking down of gravitational EFT. So far, we mainly discussed scattering amplitudes in flat space, but dispersive sum rules have also been used to explore CFTs \cite{Caron-Huot:2020adz,Carmi:2020ekr,Alday:2022uxp,Caron-Huot:2022sdy,Chang:2023szz}. In this paper, we continue these studies by confronting the twice-subtracted dispersion relations with the picture of high-energy gravitational scattering as developed by ACV, see \figref{fig:ACV}.

The plan of the paper is as follows: In \secref{sec:gavPoleandDR}, we explore explicit models for gravitational scattering combined with the twice-subtracted dispersion relations and see how the graviton pole is generated. In \secref{sec:gravPoleSR}, we construct a family of positive functionals $\psi_{a}(q)$ for which the discontinuity in the smeared version of \eqref{eq:dispersivepole} is nonnegative. We then use these functionals to derive the Tauberian theorem for the graviton pole and explore its implications. Overall, we find that the picture advocated by ACV is consistent with twice-subtracted dispersion relations. 
In \secref{sec:PoleInAdSCFT}, we generalize the analysis of gravitational scattering to CFTs. We show that the eikonal and stringy scattering models in AdS correctly generate the stress-tensor pole in the Lorentzian inversion formula. We also comment on the graviton pole generation in free theories and non-holographic CFTs. We end the paper with conclusions and a discussion of open problems.

\section{The graviton exchange and dispersion relations}\label{sec:gavPoleandDR}

In this section, we review several mechanisms for generating the tree-level gravitational exchange via dispersion relations. The three examples we consider are tree-level scattering in string theory, eikonal scattering in general relativity, and eikonal scattering in Kaluza-Klein theory. Let us recall that the graviton exchange for the elastic scattering process $A,B\to A,B$ of nonidentical scalar particles takes the form \cite{DiVecchia:2023frv}
\be
\label{eq:tree-level}
T^{\text{exchange}}(s,t) = - {8 \pi G_N \over t} \frac{1}{2} \left((s-m_A^2-m_B^2)^2 + (u-m_A^2-m_B^2)^2 -t^2  -\frac{8 m_A^2 m_B^2}{d-2} \right) \ .
\ee
This process is represented in \figref{fig:treelevel}, and by taking the small $t$ limit, we recover \eqref{eq:assymptotic_t0}. For simplicity, we set $m_A=m_B=m$ below.

Our starting point is the twice-subtracted dispersion relations for this elastic amplitude, which we can write as follows   \cite{Roy:1971tc,Guerrieri:2021tak}
\be
\label{eq:disprel}
T(s,t) = g(t) + \frac{1}{\pi}\int\limits_{0}^{\infty} ds'  \frac{T_s(s',t)}{s'^2} \(\frac{s^2}{s'-s} + \frac{u^2}{s'-u}\) ,
\ee
where $s = -(p_1+p_2)^2$, $u = -(p_1+p_3)^2$, $t=-(p_1+p_4)^2$ and $s+t+u=4m^2$.\footnote{In writing \eqref{eq:disprel} we assumed maximal analyticity. It is easy to relax this assumption by using low-energy arcs \cite{Bellazzini:2020cot}. It does not change any of the conclusions, and for simplicity, we use here the amplitude $T(s,t)$ itself.} We have defined the discontinuity of the amplitude as
\be
T_s(s,t) \equiv \lim_{\eps \to 0} {T(s+ i \epsilon,t) - T(s-i \epsilon,t)\over 2 i} \ . 
\ee
In addition, $g(t)$ is the so-called subtraction term, which can be expressed in terms of the amplitude, for example, by evaluating $T(-t/2,t)$ in \eqref{eq:disprel} and solving for $g(t)$.
 
The representation \eqref{eq:disprel} is expected to hold for $t<0$ in gravitational theories \cite{Haring:2022cyf}. It makes the $s-u$ crossing symmetry manifest and guarantees that at least for fixed $t<0$ the amplitude is bounded in the Regge limit and has the correct analytic properties. We can rewrite \eqref{eq:disprel} as follows
\be
T(s,t)-g(t)-\frac{1}{\pi}\int\limits_{0}^{-t} ds'  \frac{T_s(s',t)}{s'^2} \(\frac{s^2}{s'-s} + \frac{u^2}{s'-u}\) = \frac{1}{\pi}\int\limits_{-t}^{\infty} ds'  \frac{T_s(s',t)}{s'^2} \(\frac{s^2}{s'-s} + \frac{u^2}{s'-u}\) \ . 
\ee
We can then do the low-energy expansion on both sides and match the $s^2+u^2$ term. In this way, we arrive at \eqref{eq:dispersivepole}.

\subsection{Tree-level string theory}
\label{sec:treelevelstring}

Perhaps the simplest example to consider is tree-level scattering in string theory \cite{Kawai:1985xq}. Let us consider the four-point tree-level scattering amplitude of dilatons in type II string theory\footnote{In this example, the external particles are identical and massless $m=0$.}
\begin{equation}\label{eq:stringtree}
    T(s,t) = 8\pi G_N \(\frac{tu}{s} + \frac{su}{t} + \frac{st}{u}\)\frac{\G(1-\frac{\alpha's}{4})\G(1-\frac{\alpha't}{4})\G(1-\frac{\alpha'u}{4})}{\G(1+\frac{\alpha's}{4})\G(1+\frac{\alpha't}{4})\G(1+\frac{\alpha'u}{4})}\, .
\end{equation}
At finite $G_N$, we expect this amplitude to be a good approximation at low energies. A remarkable fact about this amplitude is that it satisfies the twice-subtracted dispersion relations for $t<0$ as it behaves as $T(s,t)\sim s^{2+\frac{\alpha' t}{2}}$ in the Regge limit $s \to \infty$, $t$ fixed. Next, let us explicitly see how the graviton exchange is generated in the dispersion relations.

We focus on the low-energy limit of the amplitude, or in other words, we consider $\alpha't, \alpha' s \ll 1$. The dispersion relation then becomes
\be
T(s,t) =8 \pi G_N \left(-{t^2 \over s}-{t^2 \over u}-{3 \over 2}t \right) +O(\alpha' s,\alpha' t) + \frac{1}{\pi}\int\limits_{{4 \over \alpha'}}^{\infty} ds'  \frac{T_s(s',t)}{s'^2} \(\frac{s^2}{s'-s} + \frac{u^2}{s'-u}\) ,
\ee
where we see that the $s$- and $u$-channel poles have been easily reproduced from the discontinuity $T_s(0,t)$. Comparing this to the full amplitude, we see that
\be
-{8 \pi G_N \over t} {s^2 + u^2 \over 2} + O(\alpha' s, \alpha' t) = \frac{1}{\pi}\int\limits_{{4 \over \alpha'}}^{\infty} ds'  \frac{T_s(s',t)}{s'^2} \(\frac{s^2}{s'-s} + \frac{u^2}{s'-u}\) \ . 
\ee
The LHS is reproduced by the high-energy limit $\alpha' s' \gg 1$ of the dispersive integral. In this way, we get
\begin{equation}
 \label{eq:dispint}
-{8 \pi G_N \over t} + O(\alpha' t) = 2 \int_{{4 \over \alpha'}}^\infty \frac{d s'}{\pi} {T_s(s',t) \over s'^3}.   
\end{equation}
The Regge limit of the amplitude $s \to \infty$, $t$ fixed can be computed explicitly and leads to\footnote{In this paper $A(s)\simeq B(s), ~ s\to \infty$ means $\lim_{s \to \infty}{A(s) \over B(s)} = 1$.}
\begin{align}
\label{eq:Reggestring0}
    T(s,t) &\simeq \frac{8\pi G_N s^2}{-t} \(\frac{\alpha's}{4}\)^{\frac{\alpha' t}{2}} \frac{\G(1-\frac{\alpha't}{4})}{\G(1+\frac{\alpha't}{4})} e^{-i \pi \frac{\alpha't}{4}} \,.
\end{align}
It is, of course, a trivial exercise to plug \eqref{eq:Reggestring0} into the dispersive integral \eqref{eq:dispint}. First, we take the discontinuity of the amplitude at small $\alpha' t \ll 1$
\begin{equation}
 \label{eq:stringyRegge}
T_s(s',t) = 8 \pi G_N {\pi \over 4 \alpha'} (\alpha' s')^{2+{\alpha' t \over 2}}.    
\end{equation}
Doing the integral over energies, we get
\be\label{eq:intDispTsStrint}
2 \int_{{4 \over \alpha'}}^\infty \frac{d s'}{\pi} {T_s(s',t) \over s'^3} =- {8 \pi G_N \over t} \int_0^\infty d x' e^{-x'} + O(\alpha' t) = -{8 \pi G_N \over t} + O(\alpha' t) , 
\ee
where we introduced the dimensionless integration variable $x' = -{1 \over 2} \alpha' t \log {\alpha' s \over 4}$. It is clear from the computation above that energies $s_*$ at which the tree-level pole is generated correspond to $x' = O(1)$, or, in other words, $-{1 \over 2} \alpha' t \log {\alpha' s_*\over 4} \sim O(1)$. 

Next, let us repeat the discussion in the impact parameter space. To do it, we write the impact parameter representation of the amplitude in the Regge limit \eqref{eq:Reggestring0}
\be
\label{eq:stringyimpactpar}
T(s,-q^2) = 2s (2\pi)^{\frac{d-2}{2}}\int_0^\infty db b^{d-3} (qb)^{\frac{4-d}{2}}J_{\frac{d-4}{2}}(bq) 2 \delta_{\rm string-tree}(s,b) , 
\ee
where the definition of the phase shift $\delta(s,b)$ and the impact parameter transform can be found in \appref{sec:convention}.

The result for the phase shift $\delta_{\rm string-tree}(s,b)$ takes the form \cite{DiVecchia:2023frv}
\begin{align}
    2 \delta_{\rm string-tree}(s,b) 
    &={\Gamma(1 - {\alpha' \over 4} \nabla_b^2) \over \Gamma(1+{\alpha' \over 4} \nabla_b^2)} \(\G\(\frac{d-4}{2}\) - \G\(\frac{d-4}{2}, \frac{b^2}{Y(s)}\)\)\frac{1}{\pi^{\frac{d-4}{2}}} \frac{G_N s}{b^{d-4}}\, ,  
\end{align}
where $Y(s) = 2 \alpha' \log {\alpha' s \over 4} - i \pi \alpha'$. 
If we consider scattering at large impact parameters $b^2\gg Y(s)$, we recover the familiar tree-level phase shift with the leading correction taking the form, see e.g. \cite{Amati:1988tn},
\begin{equation}
     2 \delta_{\rm string-tree}(s,b) \simeq 2 \delta_{\rm tree}(s,b) - \pi^{2-\frac{d}{2}} \frac{G_N s}{b^{d-4}} e^{-{b^2\over Y(s)}}\(\frac{b^2}{Y(s)}\)^{\frac{d}{2}-3}\,, ~~~ b\gg Y(s)\,.
\end{equation}
On the other hand, at small impact parameters $b\ll Y(s)$ we get 
\begin{equation}
    2 \delta_{\rm string-tree}(s,b) \simeq \frac{2 G_N s}{(\pi Y(s))^{\frac{d-4}{2}}}\(\frac{1}{d-4} - \frac{1}{d-2}\frac{b^2}{Y(s)} +\dots\)\,, ~~~ b\ll Y(s) .
\end{equation}
Notice that the phase shift develops an imaginary part related to the production of string resonances in the $s$-channel. To check the graviton pole sum rule \eqref{eq:dispersivepole}, we need to compute the following moment of the phase shift
\be
\label{eq:strTsr}
\int_{s_0}^\infty {d s \over s^2} {\rm Im} \Big( 2 \delta_{\rm string-tree}(s,b) \Big) \simeq  \frac{1}{2} \pi ^{3-\frac{d}{2}} \Gamma \left(\frac{d-4}{2}\right) {G_N \over b^{d-4}} ,
\ee
where $s_0$ is an arbitrary low-energy cutoff, and in the RHS, we only focus on the leading contribution at large $b$. We used $\Im 2 \delta_{\rm string-tree}(s,b)$ because it corresponds to taking the discontinuity $T_s(s,t)$ in \eqref{eq:stringyimpactpar}.
The dominant energies in the integral are given by $b^2 \sim Y(s)$ or, equivalently, we get for the graviton pole generation scale
\be
\label{eq:gravpolestringy}
{\alpha' \log {\alpha' s_* \over 4} \over b^2} = O(1) \ . 
\ee
Finally, doing back the Fourier transform, we correctly reproduce the pole as we already checked in \eqref{eq:intDispTsStrint}. Physically, the states responsible for generating the $t$-channel pole in the dispersion relations are the $s$-channel higher spin resonances. Let us notice that a variation of this mechanism can produce $O(G_N)$ corrections due to integrating out loops of light particles \cite{Caron-Huot:2024tsk}.

It is interesting to ask: is it possible that this simple mechanism of producing the graviton $t$-channel pole through the power-like asymptotic (or a similarly isolated singularity in the complex spin $J$-plane) can be correct in the full finite $G_N$ theory? The answer to this question is \emph{no}, as we show in \appref{sec:nonpretUnitRegge}. The reason is that this scenario is not consistent with nonperturbative unitarity. Of course, the behavior \eqref{eq:stringyRegge} is fine at intermediate energies. We can also understand this as follows: as we send $\alpha' t \to 0$, the characteristic energy $s_*$ in the dispersive integral goes to infinity, and the tree-level approximation is no longer valid.

\subsection{Eikonal scattering}
Next, we consider the model where the graviton pole is generated by gravitational loops. The model is best formulated in the impact parameter space, where  $\vec b$ is the Fourier dual of the $(d-2)$-dimensional transferred momentum $\vec q$ in the Breit frame, in which the colliding particles exchange momentum in the transverse directions and move unperturbed in the longitudinal direction, see e.g. \cite{DiVecchia:2023frv}. We consider the following ansatz for the discontinuity of the amplitude 
\be
\label{eq:treeleveldisc}
T_s^{\rm eik}(s,t=-q^2)&=2s (2\pi)^{\frac{d-2}{2}}\int_0^\infty db b^{d-3} (qb)^{\frac{4-d}{2}}J_{\frac{d-4}{2}}(bq) 2 \sin^2 \delta_{\rm tree}(s,b) ,
\ee
where the tree-level phase shift is given by the standard formula
\be\label{eq:deltatree}
 \delta_{\rm tree}(s,b) =  {\Gamma({d-4 \over 2}) \over 2 \pi^{(d-4)/2} } {G_N s \over b^{d-4}} \ . 
\ee
This ansatz for the discontinuity is motivated by the impact parameter representation of the amplitude, see \eqref{eq:impactParDef}.
The phase shift $\delta_{\rm tree}(s,b)$ encodes the Shapiro time delay experienced by a particle that crosses a gravitational shockwave \cite{Aichelburg:1970dh}. It is expected that this phase shift correctly captures the high-energy large impact parameter scattering in gravitational theories \cite{tHooft:1987vrq}. 

At low energies the discontinuity of the amplitude $T_s(s,t)$ starts as $G_N^2$ because $\sin^2 \delta_{\rm tree}(s,b)$ $\sim \delta_{\rm tree}^2(s,b) \sim G_N^2$. This fact immediately teaches us that if we want to reproduce $T(s,t)$ up to order $G_N^2$ from the dispersive integral, then all energies will contribute. However, for the term $O(G_N)$, the situation is simpler, and only high energies matter. The relevant integral takes the form
\begin{equation}
    \label{eq:dispersiveeik}
T(s,t) \simeq g(t) + {s^2 + u^2 \over 2}\int\limits_{0}^{\infty} {ds' \over \pi}  \frac{2 T_s^{\rm eik}(s',t)}{s'^3} \ .
\end{equation}
To see how the graviton pole is generated, we first perform the integral over energies
\begin{align}
\int\limits_{0}^\infty \frac{ds'}{\pi} \frac{2 \sin^2\(\frac{\Gamma\(\frac{d-4}{2}\)}{2 \pi^{\frac{d-4}{2}}}\frac{G_N s'}{b^{d-4}}\)}{s'^2} &= 2 \frac{\Gamma\(\frac{d-4}{2}\)}{2 \pi^{\frac{d-4}{2}}}G_N b^{4-d}\int\limits_{0}^\infty \frac{d \delta }{\pi} \frac{(\sin  \delta )^2 }{\delta^2} = \frac{\Gamma\(\frac{d-4}{2}\)}{2 \pi^{\frac{d-4}{2}}} G_N b^{4-d}  \ ,
\label{eq:sintegraleikonal}
\end{align}
where we changed the integration variable to $\delta = \frac{\Gamma\(\frac{d-4}{2}\)}{2 \pi^{\frac{d-4}{2}}}\frac{G_N s'}{b^{d-4}}$. Performing then the impact parameter integral in \eqref{eq:treeleveldisc}, we get for $q>0$
\be
{8 \pi G_N  \over q^2} = 4 (2\pi)^{\frac{d-2}{2}}\int_0^\infty db b^{d-3} (qb)^{\frac{4-d}{2}}J_{\frac{d-4}{2}}(bq) \left( \frac{\Gamma\(\frac{d-4}{2}\)}{2 \pi^{\frac{d-4}{2}}} G_N b^{4-d} \right) ,
\ee
where the factor of $4$ is due to a kinematical factor of $2$ in \eqref{eq:treeleveldisc} and an extra factor of $2$ in \eqref{eq:dispersiveeik} due to crossing (or equivalently the contribution of the $s$- and $u$-channels). 

We, therefore, see that the term linear in $G_N$ is correctly generated, and it takes precisely the expected form
\be
\label{eq:dispeikonal}
T(s,t) = - {8 \pi G_N \over t} {s^2+u^2 \over 2} + g(t) + ... \ . 
\ee

At this point, the fact that the graviton pole has been correctly reproduced through the dispersion relations looks somewhat miraculous and asks for an explanation. As a first remark, let us notice that purely within a semi-classical theory, we have the following true identity
\begin{equation}
    \label{eq:phaseshiftdispersion}
i (1 - e^{2 i \delta_\text{tree}(s,b)} ) = \int_0^\infty {d s' \over \pi} {2 \sin^2 \delta_\text{tree}(s',b) \over s'} \Big( {s \over s'-s} + {s \over s'+s} \Big) \ . 
\end{equation}
It can be obtained by doing the once-subtracted dispersion relations for the semi-classical phase shift ${1 - e^{2 i \delta_\text{tree}(s,b)} \over s}$.
Importantly, in deriving this identity, we used that $e^{2 i \delta_\text{tree}(s,b)}$ is analytic and decays in the upper half-plane as a function of complex $s$, which is related to the fact that particles experience the Shapiro time delay when crossing the shockwave. This fact is true in flat space and in AdS. It is also famously violated in dS at large impact parameters \cite{Gao:2000ga,Bittermann:2022hhy}.\footnote{Note that in $d=4$ flat space $\delta_\text{tree}(s,b)=- 2 G_N s \log b/L_{IR}$ and this statement only holds for $b\leq L_{IR}$, where $L_{IR}$ is the IR cutoff. If we regularize the theory by putting it in $AdS_4$, this limitation disappears since $\delta_\text{tree}^{AdS}(s,b)=2 G_N s \left(\cosh b \log \left( \coth {b \over 2} \right)  - 1 \right) \geq 0$ for any $b$ in the $R_{AdS}=1$ units. On the other hand, in $dS_4$ we get $\delta_\text{tree}^{dS}(s,b)= 2 G_N s \Big( \cos b  \log \left(\cot \frac{b}{2} \right)-1 \Big)$ which is negative when $b \sim R_{dS}=1$.} Secondly, we would like to promote \eqref{eq:phaseshiftdispersion} to an exact equation in the full quantum theory, such that the classical identity above emerges from it in the appropriate limit. Remarkably, in flat space $d>4$ and AdS, such a nonperturbative equation is known, and it is given by the twice-subtracted dispersion relations applied to the amplitude in momentum space. If we now take the dispersion relations \eqref{eq:disprel} and naively transform it to the impact parameter space considering the limit $-t \ll s$ or $s b^2 \gg 1$, we do land on the formula \eqref{eq:phaseshiftdispersion}.

From the calculation above, it is clear that for a given $b$, a characteristic energy $s_*$ at which the tree-level result is generated is given by 
\be
\label{eq:gravpoleEikonal}
\frac{\Gamma\(\frac{d-4}{2}\)}{2 \pi^{\frac{d-4}{2}}}\frac{G_N s_*}{b^{d-4}} = O(1). 
\ee
As we send $b \to \infty$, the relevant energies become arbitrarily high as expected. 

If we go back to momentum space, we get that the relevant dimensionless parameter is $\Lambda = G_N s (-t)^{{d-4 \over 2}}$. The graviton pole corresponds to taking the limit $\Lambda \to 0$; the relevant energies in the dispersive integral for generation of the graviton pole are $\Lambda = O(1)$; finally, the Regge limit corresponds to $\Lambda \to \infty$. In the Regge limit, the leading behavior of the amplitude takes the form
\be
\label{eq:Reggelimiteikonal}
\lim_{s \to \infty} T_{s}(s,t) = \frac{2^{{d \over 2}} s (-t)^{\frac{2-d}{2}}
   \left( 2 \pi^{{d-2 \over 2}} G_N s (-t)^{{d-4 \over 2}}   \Gamma
   \left(\frac{d}{2}-1\right)\right)^{\frac{d-2}{2
   (d-3)}}}{\sqrt{d-3}} \sin \left( {d-3 \over d-4} \lambda(s,t) - {\pi \over 4} (d-2) \right) ,
\ee
where $ \lambda(s,t)  = \Big( {2 \Gamma({d-2 \over 2}) \over \pi^{(d-4)/2}} G_N s (-t)^{(d-4)/2} \Big)^{1/(d-3)}$, and one can explicitly check that it does not reproduce the pole and produces a subleading contribution in the dispersive integral as we take $t \to 0$. This is in contrast to what happens in the tree-level string theory computation. 

It is also interesting to ask if we can write down a dispersive eikonal model that generates the tree-level amplitude for the scattering of identical scalar particles $A,A \to A,A$. A rather simple-minded ansatz takes the form
\be
    T_{A,A\to A,A}(s,t) &= \frac{1}{\pi}\int\limits_{0}^{\infty} ds'  \frac{T_s^\text{eik}(s',t)}{s'^2} \(\frac{s^2}{s'-s} + \frac{u^2}{s'-u} \) + \frac{1}{\pi}\int\limits_{0}^{\infty} ds'  \frac{T_s^\text{eik}(s',u)}{s'^2} \(\frac{s^2}{s'-s} + \frac{t^2}{s'-t} \) \nn \\
    &+ \frac{1}{\pi}\int\limits_{0}^{\infty} ds'  \frac{T_s^\text{eik}(s',s)}{s'^2} \(\frac{t^2}{s'-t} + \frac{u^2}{s'-u} \) . 
\ee
This amplitude appears to be fully crossing-symmetric. However, as we emphasized above, the dispersive integral 
$\frac{1}{\pi}\int\limits_{0}^{\infty} ds' \frac{T_s^\text{eik}(s',t)}{s'^2} ( \frac{s^2}{s'-s} + \frac{u^2}{s'-u} ) $ 
converges for $t<0$ only, similarly for the second term it converges for $u<0$, and the last one for $s<0$. Therefore, to use the formula above in the physical region $s+t+u = 4m^2$ requires understanding analytic continuation away from the region where the integrals converge. We have not tried to do that, and it would be very interesting to explore this model and the necessary analytic continuation in detail.

\subsection{Kaluza-Klein model}
We discuss next eikonal scattering in $(d+1)$-dimensional general relativity coupled to a real massless scalar field compactified on a circle of radius $2 \pi R$ \cite{KoemansCollado:2019lnh}.\footnote{In the language of \figref{fig:gravScaleSch}, $R\sim \ell_{KK}$ is the Kaluza-Klein scale.} We consider the scattering of massless scalars carrying zero momentum in the extra dimension. The imaginary part of the amplitude becomes  
\be\label{eq:TsKK}
T_s^\text{KK}(s,t=-q^2)&=2s (2\pi)^{\frac{d-2}{2}}\int_0^\infty db b^{d-3} (qb)^{\frac{4-d}{2}}J_{\frac{d-4}{2}}(bq) \int_0^{2 \pi R} {d b_s \over 2 \pi R} 2 \sin^2 \delta_\text{KK}(s,b,b_s) ,
\ee
where the phase shift $\delta_\text{KK}(s,b,b_s)$ is given by the following formula
\be
\delta_\text{KK}(s,b,b_s) = 2 \pi G_N^{(d+1)} {s \over (2 \pi)^{d/2}} {1 \over b^{{d-4 \over 2}} R} \sum_{n \in \mathbb{Z}} e^{i b_s n/R} \Big( {|n| \over R}\Big)^{{d-4 \over 2}} K_{{d-4 \over 2}}\Big( {|n| b \over R} \Big) \ . 
\ee
The new feature of this expression compared to the previous section is the sum over the Kaluza-Klein momentum $n/R$. At tree-level we reproduce the familiar phase shift after projecting $\delta_\text{KK}(s,b,b_s)$ to the zero Kaluza-Klain momentum
\be
\label{eq:relationx}
\int_0^{2 \pi R} {d b_s \over 2 \pi R} \delta_\text{KK}(s,b,b_s) =  \frac{\Gamma\(\frac{d-4}{2}\)}{2 \pi^{\frac{d-4}{2}}} {G_N^{(d)} s \over b^{d-4}} , 
\ee
and we have introduced the $d$-dimensional Newton constant
\begin{equation}
    G_N^{(d)}={G_N^{(d+1)} \over 2 \pi R}.
\end{equation}
It is also interesting to consider the limit of large extra dimensions $R \gg b,b_s$. We get that
\be
\delta_\text{KK}(s,b,b_s) = {\Gamma({d-3 \over 2}) \over 2 \pi^{(d-3)/2} } {G_N^{(d+1)} s \over (b^2 + b_s^2)^{{d-3 \over 2}}} ,
\ee
which is the tree-level phase shift in $d+1$ dimensions with the higher dimensional Planck scale being controlled by $G_N^{(d+1)} \sim  \ell_{Pl}^{d-1}$. Let us also recall that when $R \gg \ell_{Pl}$, the effective Planck scale for the low-energy observer appears much higher than the higher-dimensional Planck scale $\ell_{Pl}^{(d)} \ll \ell_{Pl}$, where $G_N^{(d)} \sim {G_N^{(d+1)} \over R} \sim \left( \ell_{Pl}^\text{eff} \right)^{d-2}$.

We can plug the imaginary part \eqref{eq:TsKK} into the dispersion relations \eqref{eq:dispersiveeik} and check that the graviton pole \eqref{eq:relationx} is correctly reproduced. Because $\delta_\text{KK}(s,b,b_s)\propto s$, the integral over energies is identical to \eqref{eq:sintegraleikonal}. We then trivially perform the integral over $b_s$, which is identical to \eqref{eq:relationx} and leads to the expected result.

Let us next understand the graviton pole generation scale $s_*(b)$ in this model. For $b \gg R$, we get the same formulas as in the previous section, where the graviton pole generation scale is given by the $d$-dimensional phase shift. In the opposite limit $b\ll R$, we get the following identity
\be
{2 \over \pi} \int_0^\infty {d s' \over s'^2} \int_0^\infty d \tilde b_s  \sin^2 \( {\Gamma({d-3 \over 2}) \over 2 \pi^{(d-3)/2} } {G_N^{(d+1)} s' \over b^{d-3}(1 + \tilde b_s^2)^{{d-3 \over 2}}} \) = \frac{\Gamma\(\frac{d-4}{2}\)}{2 \pi^{\frac{d-4}{2}}} {G_N^{(d+1)} \over b^{d-3} } \ , 
\ee
where we trivially canceled the $R$ dependence on both sides of the dispersion relations, and we rescaled $b_s = b \tilde b_s$. In this limit, the characteristic energy scale at which the LHS becomes of the same order as the tree-level phase shift in the RHS is controlled by the high-dimensional Planck scale
\be
\frac{\Gamma\(\frac{d-4}{2}\)}{2 \pi^{\frac{d-4}{2}}} {G_N^{(d+1)} s_* \over b^{d-3} } = O(1), ~~~ b \ll R \ . 
\ee
We, therefore, see that in this model, the perturbative expansion of the amplitude breaks down much earlier than one would naively expect by measuring $G_N^{(d)}$.

More generally, we can imagine a $D$-dimensional gravitational theory compactified on ${\cal M}_{D-d}$, with $D-d$ compact dimensions of characteristic scale $\ell_{KK} \gg \ell_{Pl}$. If we explore the graviton pole generation in this model we will find that for $b \gtrsim \ell_{KK}$ it is controlled by the $d$-dimensional eikonal scattering with the effective Newton's constant $G_N^{(d)} = {G_N^{(D)} \over {\rm Vol}_{{\cal M}_{D-d}}}$. On the other hand, for $b \lesssim \ell_{KK}$, the relevant regime becomes the one of $D$-dimensional eikonal scattering. This is the situation depicted in \figref{fig:gravScaleSch}.

\subsection{Transition between different regimes}
Given $b$, to understand which physical mechanism of the graviton pole generation is realized, we compare the corresponding graviton pole generation scales. The one that has a lower energy is realized. 

The characteristic impact parameter where the transition between the stringy and the eikonal graviton pole generation mechanisms takes place is obtained by comparing the corresponding energy $s_*(b)$ in \eqref{eq:gravpolestringy} and \eqref{eq:gravpoleEikonal}. We thus obtain
\be
\label{eq:estimateTRANS}
{e^{b^2/\ell_s^2} \over (b/\ell_s)^{d-4}} \sim \left( {\ell_s \over \ell_{Pl}} \right)^{d-2} \sim {1 \over g_s^2} \ , 
\ee
where $g_s^2 \sim \frac{G_N}{\ell_s^{d-2}}$ is the string coupling, and $\ell_s^2 =\frac{\alpha'}{4}=\frac{1}{M_s^2}$ is the square of the string length.

We conclude that the stringy mechanism of graviton pole generation is realized for $b \lesssim \ell_s \log^{1/2} \left( {1 \over g_s^2} \right)$, whereas the eikonal scattering is relevant for $b \gtrsim \ell_s \log^{1/2} \left( {1 \over g_s^2} \right)$ as depicted in \figref{fig:gravScaleSch}. 
Notice that in the stringy computation, we assumed that the tidal effects are subleading when evaluating the dispersive integral. This requires that $(G_N s_*(b) \ell_s^2)^{1/(d-2)}\ll \ell_s \log^{1/2} (\ell_s^2 s_*(b))$ at the stringy graviton pole generation scale, see e.g. Figure 2 in \cite{Amati:1988tn}. Plugging in this formula $s_*(b) \sim \ell_s^{-2} e^{b^2/\ell_s^2}$, the condition becomes ${G_N s_*(b) \ell_s^2 \over b^{d-2}} \ll 1$ and it is indeed satisfied for $b \lesssim \ell_s \log^{1/2}\left( {1 \over g_s^2} \right)$ and $g_s \ll 1$. In addition, to meaningfully isolate the graviton pole contribution in \eqref{eq:intDispTsStrint}, we need to take $b \gg \ell_s$. 

In the presence of large extra dimensions, $\ell_{KK} \gg \ell_{s}$, we can consider both $d$- and $D$-dimensional eikonal scattering, see \figref{fig:gravScaleSch}. In the latter case, we need to change $d \to D$ in the estimate of the eikonal/stringy transition point \eqref{eq:estimateTRANS} and use $g_s^2 \sim {G_N^{(D)} \over \ell_s^{D-2}}$. As a result, we again get that the stringy effects take over for $b \lesssim \ell_s \log^{1/2} \left( {1 \over g_s^2} \right)$.

If we switch to the transferred momentum, we find that the stringy  mechanism is responsible for the graviton pole generation for 
\begin{equation}
    \frac{1}{\log\left(\frac{1}{g_s^2}\right)}\lesssim -\alpha't \ll 1\, .
\end{equation}
On the other hand, eikonal scattering generates the graviton pole for $0<-\alpha' t \lesssim \frac{1}{\log\left(\frac{1}{g_s^2}\right)}$.

\subsection{Graviton pole generation versus graviton pole unitarization}

In the models above, we consider the question of \emph{the graviton pole generation} in the dispersion relations. Let us briefly discuss its relationship to \emph{the graviton pole unitarization}, which we now quickly introduce. We can expand the scattering amplitude into the partial waves
\be\label{eq:PWinMain}
T(s,t) = {1 \over 2} \sum_{J=0}^\infty n_J^{(d)} f_J(s) P_J^{(d)}(\cos \theta), ~~~ \cos \theta = 1+ {2 t \over s - 4m^2} \ ,
\ee
where our conventions can be found in \appref{sec:convention}. We can then introduce the partial waves $S_J(s)$ as follows
\begin{equation}
    S_J(s) = 1+i{(s-4m^2)^{d-3\over2}\over  \sqrt s} f_J(s) = 1+ i a_J(s)\,,
\end{equation}
where we defined $a_J(s) \equiv {(s-4m^2)^{d-3\over2}\over  \sqrt s} f_J(s)$. Unitarity is the statement that
\be
\label{eq:unitarityMain}
| S_J(s) | \leq 1 , ~~~ s  \geq 4 m^2 . 
\ee
Let us perform the partial wave projection of the graviton pole $- {8 \pi G_N s^2 \over t}$. Performing the integral, we get for $s \gg m^2$
\begin{equation}
   S_J(s) = 1 + i G_N s^{{d-2 \over 2}} c_J \,,
\end{equation}
where $c_J$ scale as $c_J\sim 1/J^{d-4}$ at large spin. 
We see that at energies of order $s \sim M_{Pl}^2$ the tree-level amplitude violates unitarity, and the quantum corrections become important.\footnote{Due to the dependence of $c_J$ on $J$, we expect that the quantum corrections become important at large spin at $s\sim M_{Pl}^2 J^{2(d-4)\over 4-2}$.}

In this section, we discussed the graviton pole generation for various models, but we have not checked that they unitarize the graviton pole at finite spin $J$. Let us briefly comment on this:
\begin{itemize}
\item We explore finite $J$ unitarity of the eikonal model in \appref{sec:AppetizerEikonalModel}. We find that it leads to unitary partial waves at all energies. We expect that the same should happen in the eikonal KK model.
\item It is easy to see the partial waves of the tree-level scattering in string theory violate unitarity at high energies \cite{Soldate:1986mk}. It is related to the fact that the simple Regge behavior $s^{j(t)}$ with $j(0)=2$ is not consistent with unitarity \eqref{eq:unitarityMain}, see \appref{sec:nonpretUnitRegge} for more details.
\item An interesting mechanism for graviton unitarization was found in \cite{Guerrieri:2021ivu,Guerrieri:2022sod}. There the low spin $J$ partial waves are unitarized by a set of resonances called \emph{graviballs} \cite{Blas:2020och,Blas:2020dyg}. It is expected that for this class of amplitudes, it is a low-spin phenomenon, and at higher spins, the usual eikonal physics is correctly reproduced.\footnote{We thank Andrea Guerrieri for discussions on the physics of these amplitudes.}
\end{itemize}
Overall, the partial waves are related to the fixed impact parameter scattering at large spin $J$ such that 
\be
J = {b \sqrt{s-4m^2} \over 2} \gg 1 .
\ee
Notice that any finite number of partial waves in the dispersion relations cannot generate the graviton pole. Therefore, the graviton pole is strictly speaking related to the large $J$ physics of the amplitude. We will see in the next section when we consider the sum rules obtained by smearing the scattering amplitude over the exchanged momentum $t=-q^2$ that, in this case, all spins become important.

\section{The graviton pole sum rules}\label{sec:gravPoleSR}

Next, we aim to generalize the analysis of the previous section to scenarios where the details of the UV completion are unknown. We assume that the twice-subtracted dispersion relations hold. However, rather than explicitly specifying the discontinuity of the amplitude $T_s(s,t)$ and verifying that it correctly reproduces the graviton pole, we derive the conditions that $T_s(s,t)$ must satisfy to achieve this.

To make progress, we search for \emph{positive functionals} $\int_0^{q_0} dq q \psi(q)$ that act on the dispersion relations in a way that ensures the contribution from UV physics above a cutoff scale $M$ is \emph{non-negative} \cite{Caron-Huot:2021rmr}.  Meanwhile, we assume that the IR part of the dispersion relations is calculable using EFT, which provides a nontrivial constraint on the unknown UV physics. Different choices of $\psi(q)$ correspond to different scattering experiments, which becomes particularly evident in impact parameter space. Specifically, the range of integration over the transferred momentum $0 \leq q \leq q_0$ translates into corresponding support in impact parameter space $b \lesssim {1 \over q_0}$.\footnote{It decays at large $b$ as a power.}

A convenient set of such positive functionals that we will use is given by
\be
\psi_{\ea} = \(\frac{q}{q_0}\)^\ea\(1-\frac{q}{q_0}\)^{(d-1)/2}, ~~~ 0<\ea\leq d-4 \ . 
\ee
Considering the limit $\ea \to 0$, we find that the IR part of the amplitude is dominated by the graviton pole $\lim_{\ea \to 0} T_{\psi_{\ea}}^{IR} \sim {G_N \over \ea}$. We then apply the Hardy-Littlewood Tauberian theorem, see \appref{app:tauberian}, to derive a universal prediction for the behavior of integrated imaginary parts of the partial waves at high spin, which guarantees that the graviton pole is correctly reproduced. 

An interesting situation arises when the low-energy amplitude is known and is dominated by gravitational interactions (e.g. dilaton scattering in string theory or graviton scattering in general). In this case, we can use a large family of positive functionals and sum rules to constrain the UV completion of gravity.

We can use positive functionals to test the physical picture of gravitational scattering against the basic physical principles of analyticity, unitarity, and crossing that led to the twice-subtracted dispersion relations. For example, given a functional $\psi$ and the energy scale $s_*(\psi)$ at which the tree-level amplitude is generated, we get an upper bound on the scattering amplitude at energies $s>s_*(\psi)$. This fact allows us to explore scattering in strongly coupled regions where current computational techniques are not under good control. We conclude that the existing physical picture of transplanckian scattering and the twice-subtracted dispersion relations are perfectly consistent.

\subsection{Positive functionals}\label{sec:PosDR}
The analysis in the case where the precise form of the UV physics is not known relies on having a positive functional. 
This idea was used already, for example, in \cite{Caron-Huot:2021rmr,Haring:2022cyf}, and here we will use it explicitly 
at small $\ea$ as this is required to reproduce the pole.

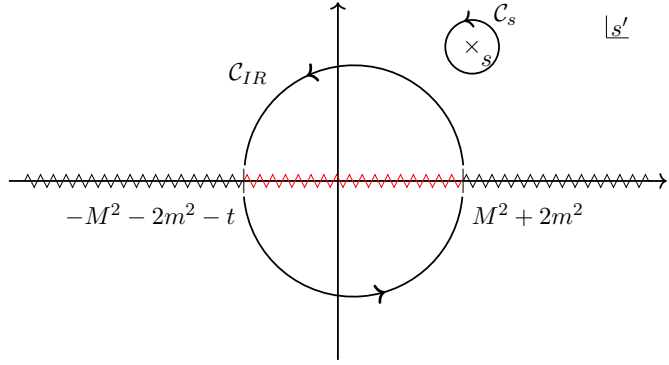
\begin{figure}[h!]
	\centering
	\scalebox{0.85}{	\begin{tikzpicture}[scale=0.7]
	\tikzmath{\x1 = 7; \y1 =4; 
		\M = \x1/2.5 ; \t = -\M/4;
		\m=3/5;
		\arcofset=0.25;}

	\coordinate (axisR) at (\x1*1.05,0);
	\coordinate (axisL) at (-\x1*1.05,0);
	
	\coordinate (axisU) at (0,\y1);
	\coordinate (axisD) at (0,-\y1);
	
	\coordinate (spole) at (3,3);
	\coordinate (pole1) at (-\t, 0);
	
	\coordinate (poleGs) at (0,0);
	\coordinate (poleGu) at (2*\m -\t,0);
	
	\coordinate (polems) at (\m,0);
	\coordinate (polemu) at (3*\m-\t,0);
	
	\draw[->, thick] (axisL)-- (axisR);
	\draw[->, thick] (axisD)--(axisU);
	
	\draw (\x1-1,\y1-1+0.2)--(\x1-0.5,\y1-1 +0.2);
	\draw (\x1-1,\y1-1 +0.2)--(\x1-1 ,\y1-1 +0.7);
	\draw (\x1-1 +0.3,\y1-1 +0.45) node{$s'$};
	
	\draw (\M,0) node[below=15pt, right] {$\Mgap^2 + 2m^2$} node{$|$};
	\draw (-\M-\t,0) node[below=15pt, left] {$-\Mgap^2-2m^2-t$} node{$|$};
	\draw[black, decoration = {zigzag, segment length = 2mm, amplitude = 1mm}, decorate] (\M,0) -- (\x1,0) ;

	\draw[black, decoration = {zigzag,segment length = 2mm, amplitude = 1mm}, decorate] (-\x1,0) -- (-\M-\t,0);

	\draw[red, decoration = {zigzag,segment length = 2mm, amplitude = 1mm}, decorate] (-\M-\t,0) -- (\M,0);

	\draw (spole) node[below right] {$s$} node{$\times$};
	\draw[thick,-, decoration={ markings,mark = at position 0.3 with {\arrow[line width=1pt]{>}}}, postaction={decorate}]  		(spole)  circle (0.6);
	\draw (spole) node[above right=6pt] {${\cal C}_s$};

	\draw (-2,2.4) node {${\cal C}_{IR}$};
	
	
	
	

	\draw[thick,-, decoration={ markings,mark = at position 0.65 with {\arrow[line width=1.2pt]{>}}}, postaction={decorate}]  (\M,\arcofset+0.1) arc (5:175:\M +\t/2);
	{\tiny }

	\draw[thick,-, decoration={ markings,mark = at position 0.6 with {\arrow[line width=1.2pt]{>}}}, postaction={decorate}]  (-\M-\t,-\arcofset-0.1) arc (185:355:\M +\t/2);
	{\tiny }
	
\end{tikzpicture}}
\caption{Definition of $\That(s,t)$ using the contour integral.}
\label{fig:contourIR}
\end{figure}

To start, it is convenient to use a dispersion relation where the IR part of the amplitude is defined by an arc around the origin. Explicitly, we define $\That(s,t)$ by
\begin{equation}
    \That(s,t) = \left(\int_{\mathcal{C}_s} + \int_{\mathcal{C}_{IR}}\right)\frac{ds'}{2\pi i}\frac{T(s',t)}{s'-s} \frac{(s-2m^2 +t/2)^2}{(s'-2m^2 +t/2)^2}\,,
\end{equation}
where $\mathcal{C}_s$ and $\mathcal C_{IR}$ are given by the contour in \figref{fig:contourIR}. The contour integral $\mathcal{C}_s$ gives $T(s,t)$, and $\mathcal{C}_{IR}$ is the contribution from the IR. Deforming the contour, we can write it as an integral over the discontinuity in the UV and gives
\begin{equation}\label{eq:ThatUV}
    \That(s,t)= \frac{1}{\pi}\int\limits_{\Mgap^2 + 2m^2}^{\infty} ds' T_s(s',t) \frac{(s-2m^2 +t/2)^2}{(s'-2m^2 +t/2)^2} \(\frac{1}{s'-s} + \frac{1}{s'-u}\)\,,
\end{equation}
where the contour at infinity has been dropped as the amplitude decays faster than $s^2$ at infinity \cite{Haring:2022cyf}.

\begin{figure}[htbp]
    \centering
    \includegraphics[width=0.6\textwidth]{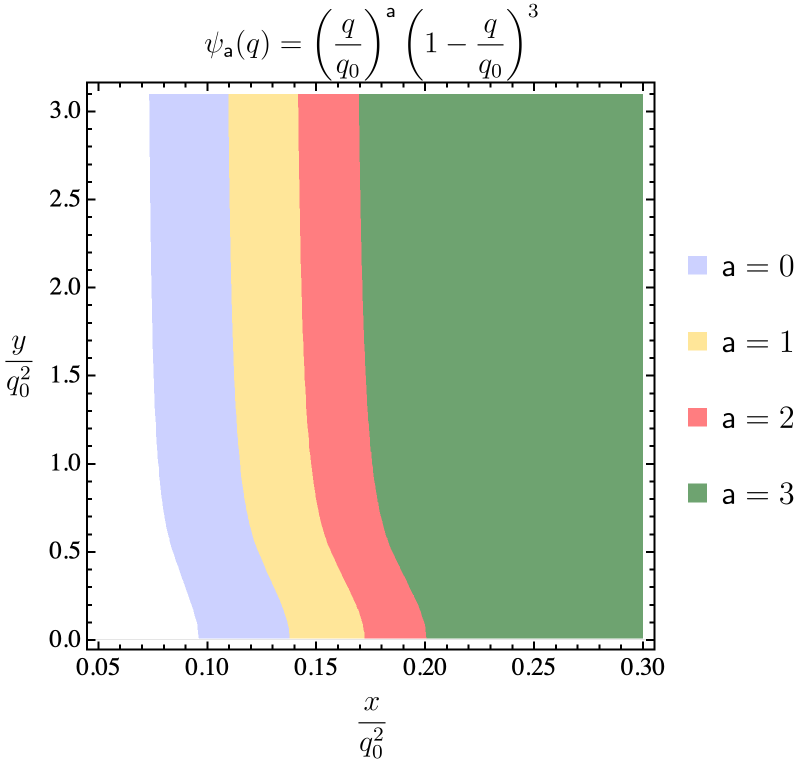}
    \caption{Region where the kernel $\cC[\psi](x,y;J,x')$ is positive with the functional $\psi_{\ea}(q) = q_0^{-\ea-\frac{d-1}{2}} q^\ea (q_0-q)^{d-1\over 2}$ in $d=7$. Here we show the result for distinct values of $\ea$, and we chose $m\ll \Mgap=q_0$. }
    \label{fig:RegionPositive}
\end{figure}

Then, we apply the smearing \eqref{eq:SmearedDef} over the dispersion relation. It is convenient to define the complex variable 
\begin{equation}
    s(x,y) = 2m^2 + (x+iy)\,,
\end{equation}
and take the imaginary part. After a series of steps, we obtain 
\begin{equation}\label{eq:ImThatDR}
    \begin{split}
        \Im \left[\That_\psi(s(x,y))\right]&= {1\over\pi}\int_0^{q_0}dq\ q\psi(q)\ \int\limits_{\Mgap^2}^\infty dx' \ T_s(s(x',0),-q^2)R(x,y; x', -q^2)  \, ,
\end{split}
\end{equation}
where we defined
\begin{equation}
    R(x,y; x', -q^2) = \frac{(2x-q^2)(2x'-q^2)y}{[y^2+(x-x')^2][y^2+(x+x'-q^2)^2]} \, . 
\end{equation}
The last step is to decompose the discontinuity of the amplitude $T_s(s,-q^2)$ in partial waves \eqref{eq:PWinMain}, which leads to
\begin{equation}\label{eq:SumRuleImTpsi}
    \Im[\widehat T_\psi(s(x,y))]={1\over\pi}\int\limits_{\Mgap^2}^\infty dx'\sum_{J=0}^\infty \njdnID \Im f_J(s(x',0)) \cC[\psi](x,y;J,x')\, .
\end{equation}
In the formula above we defined the smearing of the Legendre polynomials against the kernel by
\begin{equation}\label{eq:PositivityCpsi}
    \cC[\psi](x,y;J,x')\equiv\int_0^{q_0}dq\ q\psi(q) \left[ \Pjd\(1- \frac{2q^2}{x'-2m^2}\)R(x,y; x', -q^2) \right]\,.
\end{equation}
Provided unitarity is satisfied, obtaining a dispersion relation where the integrand is non-negative is now a kinematic statement;
there exists a non-trivial space of functionals $\psi(q)$ for which 
\begin{equation}\label{eq:conditionPosCpsi}
    \cC[\psi](x,y;J,x')\geq 0 ~~ \text{for}~ (x,y)\in R
\end{equation}
where $R$ is a region that depends on the chosen functional. The space of functionals can be characterized, see \appref{app:posFunctional}, and for our purposes, we will use a set of functionals parametrized by $\ea$
\begin{equation}\label{eq:psiaFamilly}
    \psi_{\ea} = \(\frac{q}{q_0}\)^\ea\(1-\frac{q}{q_0}\)^{\frac{d-1}{2}}\,, ~~~ 0<\ea \leq d-4 \ . 
\end{equation}
With this choice, for a given $\psi_\ea$, there is a region
\begin{equation}
    R_\ea:~ y\geq 0, x\geq x_*(y,\ea)\,,
\end{equation}
where \eqref{eq:conditionPosCpsi} is satisfied. We show the explicit region in \figref{fig:RegionPositive} for the case $d=7$. For the arguments in the next section, it is not so important where positivity is satisfied. The important point is that a non-empty region exists.

\subsection{Tauberian theorem for the graviton pole}\label{sec:TauberianTHMDerivation}
In the previous section, we have explicitly built a family of positive functionals $\psi_\ea(q)$ which allows to express the amplitude $\That_{\psi_{\ea}}(s)$ in terms of a non-negative integral in the UV \eqref{eq:ImThatDR}. In this section, we will use this family and take $\ea\to 0$ to zoom on the graviton pole.

Notice that the subtracted amplitude has the graviton pole asymptotic unchanged as $\ea\to 0$
\begin{equation}\label{eq:1overapoleIm}
        \Im[\widehat T_{\psi_{\ea}}(s(x,y))] \overset{\ea\to0}{=} \frac{16 \pi G_N xy}{\ea}  + \cO(\ea^0)\,.
\end{equation}
This is equivalent to saying that any finite energy part of the dispersive cut $s \in [0, \Mgap^2]$ cannot reproduce the graviton pole, or in other words, the graviton pole comes from the UV. The basic reason is that the kernel of the dispersive integral is regular at $\ea=0$.
For $(x,y)\in R_\ea$, we therefore get the following relation
\begin{equation}
    \frac{16 \pi G_N xy}{\ea} = \lim_{\ea \to 0}  {1\over\pi}\int_0^{q_0}dq\ q \ \psi_{\ea}(q) \int\limits_{\Mgap^2}^\infty dx' \ T_s(s(x',0),-q^2)R(x,y; x', -q^2) \geq 0 \,, 
\end{equation}
where we chose $\psi_\ea$ given by \eqref{eq:psiaFamilly}.
This immediately implies that gravity is attractive $G_N \geq 0$ \cite{Caron-Huot:2021rmr}.
The regime of interest in the dispersive integral is large $x'$. Indeed, no finite interval can reproduce the pole. In this regime ($x'\gg x,y,q^2$) the kinematical kernel becomes\footnote{This can always be achieved by choosing $m^2\sim x\sim y\sim q_0^2 \ll \Mgap^2$.}
\begin{equation}\label{eq:kernelExpamsion}
    R(x,y;x',-q^2){=} \frac{4xy}{x'^3}(1 + \cO(q^2) + \cO(1/x'^2))\,,
\end{equation}
and the sum rule of interest becomes
\begin{equation}\label{eq:sumrule}
    \frac{4 \pi G_N}{\ea} = \lim_{\ea \to 0} \int_0^{q_0}dq\, q \, \psi_{\ea}(q)  {1\over\pi}\int\limits_{\Mgap^2}^\infty {dx' \over (x')^3} \ T_s(s(x',0),-q^2) .    
\end{equation}

Next, we expand the discontinuity of the amplitude in partial wave \eqref{eq:PWinMain}
\begin{equation}\label{eq:PoleSRinPW}
 \frac{4 \pi G_N}{\ea} = \lim_{\ea \to 0}  {1\over\pi}\int\limits_{\Mgap^2}^\infty {dx' \over (x')^3} \sum_{J=0}^\infty  \njdnID (x')^{4-d\over 2}\Im a_J(s(x',0)) \[ \int_0^{q_0}dq\ q \psi_{\ea}(q)  \Pjd\(1- \frac{2q^2}{x'-2m^2}\)\] .
\end{equation}
The $q$ integral can be taken and leads to 
\begin{equation}\label{eq:intpsiPj}
    \int_0^{q_0}dq\ q \psi_{\ea}(q)  \Pjd\(1- \frac{2q^2}{x'-2m^2}\)\simeq q_0^2\frac{\, _1F_2\left(\frac{\ea}{2}+1; {d\over 2}-1 ,\frac{\ea}{2}+2;-\frac{J^2 q_0^2}{x'-2m^2}\right)}{\ea+2}\,,
\end{equation}
where the correction are suppressed in $\cO(1/x')$ and $\cO(1/J)$. Finite $x', J$, cannot reproduce the pole, only the large $x', J$ regime is relevant in the sum rules. Furthermore, the corrections are regular as $\ea\to 0$, and, therefore, can be dropped.
Plugging this expression back into \eqref{eq:PoleSRinPW}, it is convenient to introduce the impact parameter representation for partial waves
\begin{equation}
    1= \int_0^\infty db\,  \delta\(b-\frac{2J}{\sqrt{x'-2m^2}}\)\,, 
\end{equation}
and \eqref{eq:PoleSRinPW} now becomes\footnote{Here, we commuted integral over $b$ with the dispersive integral and the sum over spins. This step is justified as \eqref{eq:PoleSRinPW} converges absolutely, and the Fubini theorem can be applied.}
\begin{align}
    \frac{4 \pi G_N}{\ea} &= \lim_{\ea \to 0} \int_0^\infty db\, q_0^2 \frac{\, _1F_2\left(\frac{\ea}{2}+1;\frac{d}{2}-1,\frac{\ea}{2}+2;-\frac{b^2 q_0^2}{4}\right)}{\ea+2}   \nn \\
    &\quad \times{1\over\pi}\int\limits_{\Mgap^2}^\infty {dx' \over (x')^3}  \sum_{J=0}^\infty  \njdnID \delta\(b-\frac{2J}{\sqrt{x'-2m^2}}\) (x')^{4-d\over 2}\Im a_J(s(x',0))\,.
\end{align}
Introducing now the positive moments of the partial wave amplitudes which appear in \eqref{eq:PoleSRinPW}
\begin{equation}
\xavg{\Im a(b)}_k \equiv \frac{1}{\pi} \int\limits_{\Mgap^2}^\infty {dx' \over (x')^{k}}\sum_{J=0}^\infty\delta\(J-\frac{b\sqrt{x'-2m^2}}{2}\) \Im a_J(s(x',0)) \geq 0\,,
\end{equation}
and using that the pole can only be reproduced at large $b$, we obtain 
\begin{equation}
  \lim_{a \to 0} \int_0^\infty db  b^{d-5-a} \xavg{\Im a(b)}_2 = \(\frac{1}{2} \pi^{2-\frac{d}{ 2}} \Gamma \left(\frac{d-4}{2}\right) G_N\) \frac{1}{\ea} \,, ~~ \ea>0
\end{equation}
We can next use the Hardy-Littlewood Tauberian theorem for Laplace transform, see \appref{app:tauberian}, to get
\begin{equation}\label{eq:tauberianImpactb}
    \int_0^{\bar b} db\, b^{d-4} \xavg{\Im a(b)}_2 = \(\frac{1}{2} \pi ^{2-\frac{d}{2}} \Gamma \left(\frac{d-4}{2}\right) G_N\)\bar b + \dots ~~~,
\end{equation}
where $...$ are subleading at large $\bar b$. This is nothing but saying that on average $\xavg{\Im a (b)}_2\sim \(\frac{1}{2} \pi ^{2-\frac{d}{2}} \Gamma \left(\frac{d-4}{2}\right) G_N\) b^{4-d}$ at large $b$. 

Let us comment on the definition of the impact parameter `partial waves' obtained in this derivation. We defined them by
\begin{equation}
    a(s,b) = \sum_{J=0}^\infty\delta\(J-\frac{b\sqrt{s-4m^2}}{2}\)  a_J(s)\,,
\end{equation}
and therefore they have to be understood in a distributional sense. However, upon slightly averaging in $b$ around the value $J=\frac{b\sqrt{s-4m^2}}{2}$ we recover the usual partial waves. Alternatively, at large $s$, fixed $b$, the sum can be approximated by an integral, and we again get that $a(s,b)= a_{J=\frac{b\sqrt{s-4m^2}}{2}}(s)+ \dots$.

\paragraph{Partial waves in $J$-space}\label{sec:taubJdiscussion}
\mbox{}\\
In the argument we effectively approximated the hypergeometric function in \eqref{eq:intpsiPj} by its asymptotic $\frac{J^2 q_0^2}{x'}\gg1$
\be
q_0^2\frac{\, _1F_2\left(\frac{\ea}{2}+1; {d\over 2}-1 ,\frac{\ea}{2}+2;-\frac{J^2 q_0^2}{x'-2m^2}\right)}{\ea+2}  \simeq \frac{d-4}{4} J^{-a-2} (x')^{\frac{a}{2}+1}.
\ee
If we again introduce the partial wave moments 
\begin{equation}
\xavg{\Im a_J }_k \equiv \frac{1}{\pi} \int\limits_{\Mgap^2}^\infty {dx' \over (x')^{k}} \Im a_J(s(x',0)) \geq 0\,, 
\end{equation}
we then get the following sum rule
\begin{equation}\label{eq:finalSRinPW}
 \sum_{J =0}^\infty J^{d-5-\ea} \xavg{\Im a_J}_{{d \over 2} -\ea/2} = \left( 2^{3-d} \pi^{2-{d \over 2}} \Gamma\({d -4\over 2} \) G_N \right) {1 \over \ea} + ... , ~~~ \ea>0 \,.
\end{equation}
Now we see that compared to the discussion in the impact parameter space, we cannot derive a definite statement about $\xavg{\Im a_J}_{{d \over 2}}$ using this equation beyond the fact that at large spin it scales as $\sim G_N J^{4-d}$ with possibly a model-dependent, $J$-independent proportionality coefficient. This coefficient cancels in the ratios $\xavg{\Im a_{J+2}}_{{d \over 2}}/\xavg{\Im a_J}_{{d \over 2}}$ and we can consider the energy scale $s_\text{ho}(J)$ at which this ratio becomes $O(1)$ (and $\xavg{\Im a_J}_{{d \over 2}}$ becomes $\sim G_N J^{4-d}$). This is a signal of the higher-spin onset as defined in \cite{Caron-Huot:2024lbf}. The sum rule \eqref{eq:finalSRinPW} `almost' allows us to apply the Wiener-Ikehara tauberian theorem, see \appref{app:tauberian}. Indeed, if the partial wave moment were independent of $\ea$ in the sum rule,  the analog of \eqref{eq:tauberianImpactb} could be obtained.  One can also check that the naive application of the theorem leads to incorrect asymptotic behavior.

For the examples considered in this paper, the length scale ${1 \over \sqrt{s_\text{ho}(J)}}$ coincides with the impact parameter $b_{\star}$ at which the gravitational EFT breakdown as defined in \figref{fig:gravScaleSch}. It would be interesting to understand if this is always the case.

\subsection{Black holes in dispersion relations}
\label{sec:blackdiscBH}
Next, we examine the black disc contribution to the scattering amplitude.
We define the black disc region by the set of scattering energies and impact parameters $b$ (or spins $J$) for which the probability of elastic scattering is negligibly small $S_J(s) \simeq 0$. This means that all the partial waves in this region are $a_J(s) \simeq i$. 

Let us consider a model where the black disc regime describes scattering at energies $s \geq s_\text{bd}(b)$. An example of such a regime could be the one that corresponds to the scattering at energies such that $b \lesssim R_\text{Sch}(s)$, where the Schwarschild radius is given by
\be
\Big( R_\text{Sch}(s) \Big)^{d-3}= {16 \pi G_N \sqrt{s} \over (d-2) \Omega_{d-2}}\,, 
\ee
where $\Omega_{d-2}=\frac{2\pi^{d-1\over 2}}{\Gamma\(d-1 \over 2\)}$ is the area of the unit $(d-2)$-sphere.
In this case we find by solving $b = R_\text{Sch}(s_{bd}(b))$
\begin{equation}
    s_\text{bd}(b) \sim {b^{2(d-3)} \over G_N^2} . 
\end{equation}
Focusing on the leading contribution we get using $\Im a(s,b)=1$ for $s'\geq s_\text{bd}(b)$
\be
T^\text{bd}(s,t) &= {s^2 + u^2 \over 2} \int_0^{\infty} {d s' \over \pi} {2 T_s^\text{bd}(s',t) \over s'^3} \cr &= 4 (2\pi)^{\frac{d-2}{2}}\int_0^\infty db b^{d-3} (qb)^{\frac{4-d}{2}}J_{\frac{d-4}{2}}(bq) \int\limits_{s_\text{bd}(b)}^\infty \frac{ds'}{\pi} \frac{1}{s'^2} \ .
\ee
The integral over energies is trivial to take and we get
\be
\label{eq:bdft}
T^\text{bd}(s,t) = {s^2 + u^2 \over 2} \int d^{d-2} \vec b e^{i \vec b \cdot \vec q} {4 \over \pi} {1 \over  s_\text{bd}(|\vec b|)} \ . 
\ee
We therefore see that even though the black disc started as a finite size region for fixed energy whose Fourier transform is analytic in $q$, the effect of integration over all energies made it non-analytic in $q$!

Consider next the black disc of the Schwarzschild radius $b^{d-3} \leq c_0 R_\text{Sch}^{d-3}(s)$ this leads to 
\be
{4 \over \pi} {1 \over  s_\text{bd}(b)} =c_0^2 \frac{256  \pi ^{2-d} \Gamma \left(\frac{d-1}{2}\right)^2}{(d-2)^2} {G_N^2 \over b^{2(d-3)}} \ . 
\ee
Performing the Fourier transform we get the following contribution to the amplitude
\be
T^\text{bd}(s,t) = c_0^2 G_N^2 {s^2 + u^2 \over 2} (-t)^{{d-4 \over 2}} \frac{2^{12-d} \pi ^{1-\frac{d}{2}} \Gamma
   \left(2-\frac{d}{2}\right) \Gamma
   \left(\frac{d-1}{2}\right)^2}{(d-2)^2 \Gamma (d-3)} ,
\ee
where in even $d$ it is understood that we get the $\log(-t)$ non-analytic term. The conclusion of this exercise is that black hole production produces a one-loop effect in the amplitude.

Interestingly, the term $\sim G_N^2 {s^2 + u^2 \over 2} (-t)^{{d-4 \over 2}}$ does arise at one loop. For graviton scattering in string theory, the coefficient of this term is completely fixed \cite{Guerrieri:2021ivu,Guerrieri:2022sod}. It would be interesting to understand if there are ways to isolate and probe this term using the dispersion relations, similar to what we have done for the tree-level graviton pole. In other words, in the computation above, we have not shown that black hole production is the only way to generate such a term.

Notice also that the integral \eqref{eq:bdft} diverges at small impact parameters, which corresponds to applying the black disc model at small energies $s_\text{bd}(b) \to 0$. This is clearly unphysical, and has to be regulated using other physical effects. For example, in string theory it is expected that the minimal energy for the collapse is given by the correspondence point $s_\text{bd} = ({M_s \over g_s^2})^2$ which produces the contribution to the amplitude $\sim G_N^2 (s^2+u^2) M_s^{d-4}$ that is analytic in $t$.

\subsection{Bound on the stringy gray disc}
\label{sec:greydisc}
We return now to the string scattering at small impact parameters. In \secref{sec:treelevelstring}, we observed how the tree-level string amplitude reproduces the graviton pole through the dispersion relations, but it violates the nonperturbative unitarity. Let us briefly notice that it satisfies the Tauberian sum rule, which takes precisely the form of \eqref{eq:strTsr}.
Furthermore, if we look at \figref{fig:ACV}, we see that the tree-level string amplitude is relevant at small impact parameters. 

In the previous section, we discussed the black disc model and observed that its contribution was at order $\cO(G_N^2)$. Here, we want to focus on the string/black hole transition region of \figref{fig:ACV} and ask \emph{how gray must this region be?} 

We consider the sum rules obtained by smearing with a functional \eqref{eq:ImThatDR}. We can chose $M={\lambda M_s} $ and write
\begin{equation}
\begin{split}
     \Im \left[\That_{\psi_\ea}(s(x,y))\right] 
     & = \int_0^{q_0} dq q \psi_\ea(q)\int_{(\lambda M_s)^2}^\infty {dx'\over \pi} \ T_s(x',-q^2)R(x,y; x', -q^2)\geq 0\,,
\end{split}
\end{equation}
where the positivity is obtained by choosing a suitable functional \eqref{eq:psiaFamilly}. We chose $q_0=\kappa M_s$ with $\kappa\leq 1$, such that the functional localizes on scattering at impact parameter $b\lesssim \ell_s$.
The amplitude on the left side can be approximated by the tree-level string amplitude \eqref{eq:stringtree} up to $\cO(G_N^2)$ corrections provided that $\lambda\ll \frac{1}{g_s}$. The tree-level string amplitude admits twice-subtracted dispersion relation, and using the identity \eqref{eq:ThatUV}, it can be written as an integral over its discontinuity
\begin{equation}
    \begin{split}
       &\int_0^{\kappa M_s} dq q \psi_\ea(q)\int_{(\lambda M_s)^2}^\infty {dx'\over \pi} \ T^{\rm string-tree}_s(x',-q^2)R(x,y; x', -q^2) + \cO(G_N^2)\\
       & \qquad\qquad\qquad\qquad= \int_0^{\kappa M_s} dq q \psi_\ea(q)\int_{(\lambda M_s)^2}^\infty {dx'\over \pi} \ T_s(x',-q^2)R(x,y; x', -q^2)\geq 0\,.
    \end{split}
\end{equation}
As we are considering a scattering experiment at $b\lesssim \ell_s$, we expect that for $s\gtrsim M_s^2/g_s^4$, the discontinuity of the amplitude to be given by a black disc (see \secref{sec:blackdiscBH}) which contribution is $\cO(G_N^2)$.\footnote{We remind the reader that $g_s^2\sim \frac{G_N}{\ell_s^{d-2}}$.} The sum rule becomes 
\begin{equation}
    \begin{split}
       &\int_0^{\kappa M_s} dq q \psi_\ea(q)\int_{(\lambda M_s)^2}^\infty {dx'\over \pi} \ T^{\rm string-tree}_s(x',-q^2)R(x,y; x', -q^2) + \cO(G_N^2)\\
       & \qquad\qquad\qquad\qquad= \int_0^{\kappa M_s} dq q \psi_\ea(q)\int_{(\lambda M_s)^2}^{M_s^2\over g_s^4} {dx'\over \pi} \ T_s(x',-q^2)R(x,y; x', -q^2)\geq 0\,.
    \end{split}
\end{equation}
We can now compute the leading $G_N$ behavior of the LHS, which leads to at large $\lambda$
\begin{equation}
    \frac{8 \pi  G_N x y \Gamma \left(\frac{\ea}{2}\right) }{(\kappa^2 \log
   \lambda)^{\ea/2}}+\dots\,.
\end{equation}
This behavior cannot be reproduced by the black disc region.  To make this term as small as possible, we maximize $\ea$ while keeping positivity, thus $\ea\to d-4$, and we obtain
\begin{equation}
\begin{split}
      \frac{8 \pi  G_N x y \Gamma \left(\frac{d-4}{2}\right) }{(\kappa^2\log
   \lambda)^{d-4\over2}}&\(1+ \cO\(\frac{1}{\kappa \log^{1/2}\lambda}\)\) \\
   &=  \int_0^{\kappa M_s} dq q \psi_{d-4}(q)\int_{(\lambda M_s)^2}^{M_s^2\over g_s^4} {dx'\over \pi} \ T_s(x',-q^2)R(x,y; x', -q^2)\geq 0\,. 
\end{split}
\end{equation}
This implies that the integral on the RHS must be of order $\cO\left({G_N/ (\log \lambda)^{d-4\over 2}}\right)$. This puts an upper bound on the amount of scattering in the region $\sqrt{s} \gtrsim {\lambda M_s}$. In fact, this also prevents having a black disc up to the strongly coupled region $\sqrt{s} \gtrsim {M_s\over g_s}$ whose contribution would be of order $\cO(G_N)$, and implies that the scattering in this region must be gray. This is in agreement with the physical picture advocated for example in \cite{Veneziano:2004er}, where they obtained 
\begin{equation}
   |1+i a(s,b)| \sim  \exp\left(-\frac{2\pi^2 G_N s \ell_s^{4-d}}{(\pi \log {s \ell_s^2})^{d-2\over 2}}\right)\,, ~~ b\lesssim \ell_s^2 \log s \ell_s^2 \ .
\end{equation}
The behavior above was derived based on the analysis of the tidal excitations of the string.
In this picture, the `grayness' of scattering at energies $\sqrt{s} \sim {M_s\over g_s}$ is controlled by $G_N$, so that ${\rm Im}\ a({M_s^2 \over g_s^2},b) \sim {1 \over (\log 1/g_s^2)^{{d-2 \over 2} }}$. We see that this scenario is perfectly compatible with the dispersion relations.

\section{The graviton pole generation in AdS/CFT}\label{sec:PoleInAdSCFT}

In this section, we discuss the graviton pole generation in the context of the AdS/CFT correspondence \cite{Maldacena:1997re,Gubser:1998bc,Witten:1998qj}.
We first consider the eikonal ansatz for the holographic CFT four-point function as described in \cite{Cornalba:2007zb,Meltzer:2019pyl}. We assume that the eikonal ansatz correctly captures the double discontinuity of the CFT four-point function in the Regge limit. We then test the consistency of this idea by plugging it into the Lorentzian inversion formula \cite{Caron-Huot:2017vep}, and verifying that the stress tensor pole and its residue fixed by the conformal Ward identities are correctly reproduced. We then repeat the same exercise for the stringy graviton pole generation mechanism. Finally, we discuss the graviton pole generation for free and non-holographic theories.

\subsection{Eikonal ansatz}

We consider the `elastic' four-point function 
$$\langle \phi(x_4) \phi(x_3) \psi(x_2) \psi(x_1) \rangle$$ of scalar primary operators with scaling dimensions $\Delta_\phi$ and $\Delta_\psi$ correspondingly. In a given CFT, this correlator is always nontrivial because in the $s$-channel OPE we get $\psi \times \psi  = 1 + T_{\mu \nu} + ...$ with the identity operator $1$ and the stress-energy tensor $T_{\mu \nu}$ always being exchanged. In the dual AdS bulk, this is related to the universality of gravitational attraction, see \figref{fig:treelevel}.

It is convenient to start with the result for the disconnected correlator
\be
\langle \phi(\infty) \phi(1) \psi(z, \bar z) \psi(0) \rangle = {1 \over (z \bar z)^{\Delta_\psi}} ,
\ee
where $\phi(\infty) = \lim_{x_4 \to \infty} x_4^{2 \Delta_\phi} \phi(x_4)$. Below, we will be interested in the Regge limit when $z, \bar z \to 0$ with ${z \over \bar z}$ fixed. We introduce the following parameterization for the cross-ratios
\be\label{eq:zzbarASsigmarho}
z = \sigma e^{\rho}, ~~~ \bar z = \sigma e^{- \rho} \ . 
\ee
The Regge limit then corresponds to $\sigma \to 0$ after analytically continuing the Euclidean correlator to the Lorentzian kinematics such that $\bar z$ goes around $1$.

In the $t$-channel, the identity operator is reproduced by summing over the double-twist operators of the schematic form $\phi \partial_{\mu_1} ... \partial_{\mu_J} \partial^{2n} \psi$. They have scaling dimension $\Delta_{n,J} = \Delta_\phi + \Delta_\psi+2n+J$ and spin $J$. It is convenient to introduce variables $h \geq \bar h \geq 0$ defined as follows
\be
\Delta = h + \bar h, ~~~~ J = h - \bar h. 
\ee
We would like to write the correlator as a sum of the double-twist operators in the generalized free field theory. To this extent it is convenient to introduce the impact parameter blocks ${\cal I}_{h, \bar h}(z, \bar z)$ in terms of which we get
\be
{1 \over (z \bar z)^{\Delta_\psi}} ={1 \over (z \bar z)^{\Delta_\psi}} \int_0^\infty d h \int_0^h d \bar h \ {\cal I}_{h, \bar h}(z, \bar z) .
\ee
The precise definition of the blocks is given via the limit of the $t$-channel blocks weighed by the generalized free field theory (GFF) OPE coefficients \cite{Fitzpatrick:2011dm}
\be
{\cal I}_{h, \bar h}(z, \bar z) \equiv \lim_{z, \bar z \to 0, h, \bar h \to \infty, z h^2, \bar z \bar h^2 - \text{fixed}}  (z \bar z)^{\Delta_\psi} \lambda^2_\text{GFF}(h, \bar h) G_{\Delta,J}(1-z, 1-\bar z) .
\ee

To write down the explicit representation for ${\cal I}_{h, \bar h}(z, \bar z)$ it is convenient to introduce a different parameterization of the cross ratios
\be
z \bar z= x^2 \bar x^2, ~~~ z + \bar z = -2 x \cdot \bar x ,
\ee
where we will consider $x^\mu$ and $\bar x^\mu$ to be future-directed time-like vectors in $\mathbb{R}^{1,d-1}$, and we use mostly plus signature. 

Conformal blocks ${\cal I}_{h, \bar h}(z, \bar z)$ admit the following representation \cite{Cornalba:2007zb}
\be
{\cal I}_{h, \bar h}(z, \bar z) &= (-x^2)^{\Delta_\phi} (-\bar x^2)^{\Delta_\psi} C(\Delta_\psi) C(\Delta_{\phi}) \int_{M^+} {d^d p \over (2 \pi)^d} {d^d \bar p \over (2 \pi)^d} \ (- p^2)^{\Delta_\phi - {d \over 2}} \ (- \bar p^2)^{\Delta_\psi - {d \over 2}} e^{p \cdot x} e^{ \bar p \cdot \bar x} \nn \\
&4 h \bar h (h^2 - \bar h^2) \delta \left( {p \cdot \bar p \over 2} + h^2 + \bar h^2 \right) \delta \left( {p^2 \bar p^2 \over 16} - h^2 \bar h^2 \right) , \label{eq:impactpar}
\ee
where $M^+$ stands for the future Milne wedge: we integrate over future-directed time-like vectors. In the formula above
\begin{equation}
    C(\Delta) = \frac{\pi ^{\frac{d}{2}+1} 2^{d-2 \Delta +1}}{\Gamma
   (\Delta ) \Gamma \left(\Delta +\frac{2-d}{2} \right)} \ . 
\end{equation}

We are now ready to write down the eikonal ansatz for the Regge limit of the $t$-channel double discontinuity of the correlator in the full interacting theory 
\be
\langle \phi(\infty) \phi(1) \psi(z, \bar z) \psi(0) \rangle = {{\cal G}(z, \bar z) \over (z \bar z)^{\Delta_\psi}} \ . 
\ee
We will only need to consider the double discontinuity of the correlator \cite{Caron-Huot:2017vep}. The eikonal ansatz for it takes the following form
\be
\label{eq:eikonalansatzCFT}
{\rm dDisc} {\cal G}_\text{eik} (z,\bar z) = \int_0^\infty d h \int_0^h d \bar h \ 2 \sin^2 \left( {\delta(S,L) \over 2} \right) \ {\cal I}_{h, \bar h}(z, \bar z) ,
\ee
where $\delta(S,L)$ is the tree-level gravitational phase shift \cite{Cornalba:2006xk,Cornalba:2006xm,Cornalba:2007zb}
\be
\delta(S,L) = 4 \pi G_N S {\cal H}_{d-1}(L) \,,
\ee
and we introduced the scattering energy $S$ and the impact parameter $L$ which are given in terms of $(h,\bar h)$ by the following formula
\be
S = 4 h \bar h, ~~~ \cosh L = {1 \over 2} \left( {h \over \bar h} + {\bar h \over h} \right) \, . 
\ee
The Newton $G_N$ constant is related to the central charge of the theory as follows
\be
c_T =\frac{d+1}{d-1}\frac{\pi^{\frac{d}{2}}\Gamma (d+1)}{\Gamma\left(\frac{d}{2}\right)^3}\frac{R_{AdS}^{d-1}}{2\pi G_N} ,
\ee
and below we always set $R_{AdS}=1$.
We also introduced the propagator in the hyperbolic impact parameter space
\be\label{eq:hyperbolicProp}
    {\cal H}_{\Delta-1}(L)=\frac{\pi^{1-\frac{d}{2}}\Gamma(\Delta-1)}{2\Gamma(\Delta-\frac{d-2}{2})}e^{-(\Delta-1)L}{}_2F_1\Big(\frac{d}{2}-1,\Delta-1,\Delta-\frac{d}{2}+1;e^{-2L}\Big).
\ee
In the Regge limit $z, \bar z \to 0$ with $z/\bar z$ fixed, which is dominated by large $S$ and large $L$ we effectively have $2 \sin^2 \left( {\delta(S,L) \over 2} \right)  \to 1$ and
\be
{\rm dDisc} {\cal G}_\text{eik} (z,\bar z) \simeq  1 \ . 
\ee

Compared to the original work on the eikonal representation of the correlation function \cite{Cornalba:2007zb} where the formula above was derived, we would like to combine it with the ACV picture of gravitational scattering that suggests that in holographic theories, the formula above should correctly capture the nonperturbative Regge limit of the correlator. 
The basic observation is that the Regge limit is controlled by the semi-classical eikonal large impact parameter physics. Assuming this, we can then perform a consistency test of this idea by plugging it into the Lorentzian inversion formula, and checking that it correctly reproduces the stress-energy pole and, more generally, if it leads to results consistent with the OPE structure of the low-energy holographic theory. 

\paragraph{Evaluation of the integrals}
\mbox{}\\
Our task is to evaluate the integrals above to find the Regge limit behavior of ${\rm dDisc} {\cal G}_\text{eik}(z,\bar z)$. It is convenient to proceed with the following set of steps. We start by using the following convenient identity for the eikonal phase $\sin^2 \left( {\delta(S,L) \over 2} \right)$
\be
2\sin^2 \left( {\delta(S,L) \over 2} \right) = - \int_{1 + \epsilon - i \infty}^{1+\epsilon+ i \infty} {d J_L \over 2 \pi i} \left( 4 \pi G_N S {\cal H}_{d-1}(L) \right)^{J_L-1} \Gamma(1-J_L) \sin {\pi J_L \over 2} ,
\ee
where $0 < \eps < 1$. The advantage of this representation is that it effectively factorizes the energy and impact parameter dependence. This representation makes it manifest that the non-perturbative Regge intercept is $J_\text{np}=1$. Indeed, in the $S \to \infty$ limit we would like to deform the contour \emph{to the left}, and the leading singularity is located at $J_L=1$.

For the impact parameter dependence, we will be interested in the contributions of large impact parameters. For this purpose, it is convenient to rewrite the power of the propagator above as follows
\begin{equation}
    \left( {\cal H}_{d-1}(L) \right)^{J_L-1} = \sum_{k=0}^\infty\alpha_k(J_L){\cal H}_{(J_L-1)(d-1)+2 k}(L) ,
\end{equation}
where at large impact parameters we have ${\cal H}_{\Delta-1}(L) \sim e^{-(\Delta-1)L}$.
The coefficients $\alpha_k(J_L)$ can be easily computed, and, for example, we have for the leading term
\be
\alpha_0(J) = \frac{2^{2-J} \pi ^{-\frac{d J}{2}+d+J-2} \Gamma
   \left(\frac{d}{2}+1\right)^{1-J} \Gamma
   (d-1)^{J-1} \Gamma \left(d
   \left(J-\frac{3}{2}\right)-J+3\right)}{\Gamma
   ((d-1) (J-1))} \ . 
\ee
We can then use the spectral representation for the propagator, see e.g. \cite{Penedones:2007ns},
\be
{\cal H}_{\Delta - 1}(L) = \int_{- \infty}^{\infty} d \nu {\Omega_{ \nu}(L) \over \nu^2 + (\Delta - {d \over 2})^2} \ , ~~~{\rm Re}[\Delta]> d/2 \,,
\ee
where $\Omega_{ \nu}(L)$ are harmonic functions in the hyperbolic space $H^{d-1}$.
To accommodate for this condition we therefore choose in the $J_L$ contour integral $1> \eps > {d \over 2(d-1)}$.

It is trivial to perform the $(h,\bar h)$ integrals using the $\delta$-functions in \eqref{eq:impactpar}. Finally, we use the following identity to perform the Fourier transforms, see e.g. \cite{Penedones:2007ns,Kulaxizi:2017ixa},
\be
 \int_{M^+} d^d p d^d \bar p {e^{x.p } e^{\bar x. \bar p } \over (-p^2)^{{d -a \over 2} } (-\bar p^2)^{{d-b \over 2} }} \Omega_{ \nu} (L) ={\pi^{d-2} \over 2^{2-a-b}} {\gamma(\nu,J_L) \gamma(- \nu,J_L)  \Omega_{\nu} (\rho) \over (-x^2 )^{a/2} (-\bar x^2  )^{b/2} }\,,
\ee
where $\rho$ was defined in \eqref{eq:zzbarASsigmarho}, and
\be
\gamma(\nu,J) &=\Gamma \left( 2 \Delta_\phi + J + i \nu - {d \over 2} \over 2 \right) \Gamma \left( 2 \Delta_\psi + J + i \nu - {d \over 2} \over 2 \right) \,,
\ee
and for us $a = 2 \Delta_\phi + J_L - 1$, $b = 2 \Delta_{\psi} + J_L - 1 $.

After these simple steps, we get the following representation for the double discontinuity of the correlator
\be
&{\rm dDisc} {\cal G}_\text{eik}(z,\bar z) =-{1 \over (2 \pi)^{2d}} C(\Delta_\phi) C(\Delta_\psi) \int_{1+\eps- i \infty}^{1+\eps+ i \infty} {d J_L \over 2 \pi i} \left( 4 \pi G_N \right)^{J_L-1} \Gamma(1-J_L) \sin {\pi J_L \over 2}\nn \\ 
& \pi ^{d-2}  4^{\Delta_\psi +\Delta \phi +J_L-2} \sum_{k=0}^{\infty } \alpha_k (J_L)   \int_{- \infty}^{\infty} d \nu { \gamma(\nu,J_L) \gamma(-\nu,J_L) \Omega_\nu(\rho) \over \nu^2 + ((J_L-1)(d-1)+{2-d \over 2} + 2k)^2} \sigma^{1-J_L} .
\ee
Foreseeing the inversion formula discussion, let us define the following integral 
\be
I(\rho,J) \equiv \int_0^1 d \sigma \ \sigma^{J-2} {\rm dDisc} {\cal G}_\text{eik}(z,\bar z). 
\ee
Using the representation above it is trivial to do the integral which produces ${1 \over J-J_L}$.

We close the $J_L$ contour \emph{to the right} and we pick the residue at $J=J_L$. The kernel has poles $J_L=3,5,...$, which we will discuss below, but for now, these are not relevant since we will be interested in $J=2$. In this way, we are left with the following expression
\be
I_{J_L=J}(\rho,J) &=-{1 \over (2 \pi)^{2d}} C(\Delta_\phi) C(\Delta_\psi) \left( 4 \pi G_N \right)^{J-1} \Gamma(1-J) \sin {\pi J \over 2}\nn \\ 
& \pi ^{d-2}  4^{\Delta_\psi +\Delta \phi +J-2} \sum_{k=0}^{\infty } \alpha_k (J)   \int_{- \infty}^{\infty} d \nu { \gamma(\nu,J) \gamma(-\nu,J) \Omega_\nu(\rho) \over \nu^2 + ((J-1)(d-1)+{2-d \over 2} + 2k)^2} .
\ee
We evaluate the $\nu$ integral focusing on the contribution of the pole in the denominator. Introducing
\begin{equation}
    \nu_k^2 + \left((J_L-1)(d-1)+{2-d \over 2} + 2k \right)^2 = 0 \ , 
\end{equation}
we get the following expression for the contribution from these poles
\be
\label{eq:sumrhoI}
I_{J_L=J}(\rho,J) &=-{1 \over (2 \pi)^{2d}} C(\Delta_\phi) C(\Delta_\psi) \left( 4 \pi G_N \right)^{J-1} \Gamma(1-J) \sin {\pi J \over 2}\nn \\ 
& \pi ^{d-2}  4^{\Delta_\psi +\Delta \phi +J-2} \sum_{k=0}^{\infty } \alpha_k (J) \gamma(\nu_k,J) \gamma(-\nu_k,J)  {\cal H}_{(J-1)(d-1)+2 k}(\rho) \\
&+(\text{double trace}), \nn 
\ee
where by $($double trace$)$ we denoted extra contributions coming from the poles of $\gamma(\nu,J)\gamma(-\nu,J)$. These will generate contributions to the OPE of double trace operators $[\phi,\phi]_{n,J}$ and $[\psi,\psi]_{n,J}$. The expression \eqref{eq:sumrhoI} is suitable for evaluation inside the inversion formula, which we do next. 

\paragraph{Lorentzian inversion formula}
\mbox{}\\
Recall that the Lorentzian inversion formula expresses the conformal partial waves in terms of the double discontinuity of the correlator. Approximating the double discontinuity by a few light operators produces the results of the light-cone bootstrap \cite{Fitzpatrick:2012yx,Komargodski:2012ek,Caron-Huot:2017vep}. In other words, we approximate the double discontinuity by the identity operator and the stress tensor in the OPE, and we ask how they are reproduced in the dual channel. Here, we ask a different question: \emph{how is the stress tensor generated by heavy operators in the dual channel?}

The structure of the Lorentzian inversion formula is such that `simple' operators like the stress-energy tensor are expected to be captured by the contribution of infinitely many operators into the OPE of the four-point function in the dual channel. It is, therefore, very interesting to see what is the OPE structure generated by the eikonal ansatz above.

A convenient formula that captures the contribution of the heavy operators to the conformal partial waves is given by (5.2) in \cite{Caron-Huot:2017vep}
\be
\label{eq:heavyinv}
c^t(\Delta,J)=C_0(\Delta,J) \int_0^1 d \sigma \sigma^{J-2} \int_{-\infty}^\infty d \rho | \sinh \rho |^{d-2} \tilde C_{\Delta+1-d}(\cosh \rho) {\rm dDisc} {\cal G}_\text{heavy}(z,\bar z) , 
\ee
where
\begin{equation}
   C_0(\Delta,J) = \frac{\Gamma (\Delta -1) \Gamma \left(\frac{J+\Delta
   }{2}\right)^4}{2 \pi ^{3/2} \Gamma
   \left(\frac{d-1}{2}\right) \Gamma \left(\Delta
   -\frac{d}{2}\right) \Gamma (J+\Delta -1) \Gamma
   (J+\Delta )} \ , 
\end{equation}
and
\begin{equation}
    \tilde C_{\Delta+1-d}(\cosh \rho) = \, _2F_1\left(d-\Delta -1,\Delta
   -1, \frac{d-1}{2},\frac{1}{2} (1-\cosh (\rho
   ))\right) \ . 
\end{equation}
Plugging $ {\rm dDisc} {\cal G}_\text{eik}(z,\bar z) $ into \eqref{eq:heavyinv}, we get
\be
c^t(\Delta,J)= 2C_0(\Delta,J) \int_{0}^\infty d \rho  (\sinh \rho)^{d-2} \tilde C_{\Delta+1-d}(\cosh \rho) I(\rho,J) \ . 
\ee

\subsection{The stress tensor residue}

For the correlator that we consider, the conformal partial waves take the form
\begin{equation}
   c^{\pm}(\Delta,J) = c^t(\Delta,J) \pm c^u(\Delta,J) =(1 \pm 1) c^t(\Delta,J) \ . 
\end{equation}
Here $\pm$ stands for the signature of operators, with $+/-$ corresponding to the continuation form even/odd spins. In our case, we have only even operators present in the $s$-channel OPE, and below to avoid clutter we omit writing the signature explicitly with the $+$ signature being implicitly understood.

They exhibit poles at the locations of the exchanged operators with the residues given by the three-point functions
\begin{equation}
    c(\Delta,J) \simeq -{\lambda_{\phi \phi {\cal O}} \lambda_{\psi \psi {\cal O}} \over \Delta - \Delta_{{\cal O}}} \ .
\end{equation}
Therefore we need to interpret the poles of the previous section in terms of local operators.

We start with the special case of $J=2$. In this case, we have a pole at $\Delta = d$ which should correspond to the stress-energy tensor for which we have in our conventions \cite{Dolan:2000ut}
\be
\label{eq:expectedres}
\lambda_{\phi \phi T} \lambda_{\psi \psi T} = {d^2 \over 4 (d-1)^2} {\Delta_\phi \Delta_\psi \over c_T} .
\ee

In our computation above, an obvious simplification in the case of ${I}_{J_L=2}(\rho,2)$ is that $\alpha_{k>0}(2)=0$ and therefore the relevant integral takes the form
\be
\label{eq:stressintegral}
\int_{0}^\infty d \rho  (\sinh \rho)^{d-2} \tilde C_{\Delta+1-d}(\cosh \rho) {\cal H}_{d-1}(\rho) = \frac{\pi ^{\frac{1}{2}-\frac{d}{2}} \Gamma
   \left(\frac{d-1}{2}\right)}{2 \Delta  (d-\Delta
   )} .
\ee
Remarkably, this integral only has a pole at $\Delta = d$ as expected.\footnote{By virtue of the shadow $\Delta \to d - \Delta$ symmetry of $\tilde C_{\Delta+1-d}(\cosh \rho)$ it also has a shadow image of the stress tensor pole.} Therefore, we have correctly reproduced the stress tensor pole in the $J=2$ sector.

Next, let us check that the stress tensor residue is correctly reproduced. 
At this point, we are ready to collect all the factors. The origin of ${1 \over c_T}$ is already clear from the formula above, which contains the $G_N^{J-1}$ factor. We also get
\be
4^{\Delta
   \psi +\Delta \phi} \gamma(-i d/2,2)\gamma(+i d/2,2) C(\Delta_\phi) C(\Delta_\psi) = 4^{1+d} \pi^{d+2} \Delta_\phi \Delta_\psi ,
\ee
which correctly captures the dependence on $\Delta_\phi$ and $\Delta_\psi$. Finally, we can verify the $d$ dependence and the overall coefficient. Plugging all the factors we find that the expected result \eqref{eq:expectedres} is exactly reproduced! To make it more explicit we can write it as follows
\be
&- 4 C_0(d,2) (4 \pi G_N) (4^{1+d} \pi^{d+2} \Delta_\phi \Delta_\psi) {\pi^{d-2} \over (2 \pi)^{d}} \left( -{\pi \over 2} \alpha_0(2) \right) \frac{\pi ^{\frac{1}{2}-\frac{d}{2}} \Gamma
   \left(\frac{d-1}{2}\right)}{2 d} = {d^2 \over 4 (d-1)^2} {\Delta_\phi \Delta_\psi \over c_T} , 
\ee
where $\frac{\pi ^{\frac{1}{2}-\frac{d}{2}} \Gamma
   \left(\frac{d-1}{2}\right)}{2 d} {-1 \over \Delta - d}$ comes from the (minus) residue of the integral \eqref{eq:stressintegral} at $\Delta = d$. We also used that $\lim_{J \to 2} \Gamma(1-J) \sin {\pi J \over 2} = - {\pi \over 2}$.

Drawing an analogy with the previous section it should also be possible to repeat the computation presented here directly at the level of the CFT dispersion relations \cite{Carmi:2019cub,Penedones:2019tng}. We found it technically easier to work directly with the Lorentzian inversion formula.  

\paragraph{Regge poles and Regge bumps}
\mbox{}\\
In the computation above if we move the $J_L$ contour further to the right, we will also pick the contributions from the poles at $J_L =2n+1,$ $n \in \mathbb{Z}_+$. They contribute to $c(\Delta,J)$ as ${1 \over c_T^{2n}}{1 \over J-(2n+1)}$, and simply correspond to the perturbative expansion of the eikonal correlator. These Regge poles were explicitly related to the analytic continuation of the perturbative OPE data in \cite{Fitzpatrick:2019efk}.

In the full nonperturbative theory, (namely, if we first set $J=3,5,7,...$ and only then expand in $1/c_T$), these Regge poles are not present and become instead `bumps' of the size ${\log c_T \over c_T^{2n}}$ as can be seen from the eikonal Regge ansatz formula. The enhanced contributions can appear in the observables that involve light-ray operators of odd spins. For example, this situation was recently discussed in the case of the energy-energy correlator \cite{Chen:2024iuv}, which is controlled by the spin $J=3$ light-ray operators \cite{Hofman:2008ar,Kologlu:2019mfz,Chang:2020qpj}.

\paragraph{Higher spins}
\mbox{}\\
Let us next briefly discuss higher spin operators with $J>2$.  
In this case we get that all $\alpha_k(J) \neq 0$, and the relevant integral becomes
\be
\label{eq:generalint}
\int_{0}^\infty d \rho  (\sinh \rho)^{d-2} \tilde C_{\Delta+1-d}(\cosh \rho) {\cal H}_{\tilde \Delta-1}(\rho) = \frac{\pi ^{\frac{1}{2}-\frac{d}{2}} \Gamma
   \left(\frac{d-1}{2}\right)}{2 (\Delta-\tilde \Delta)  (d-\Delta-\tilde \Delta)}.
\ee

However, for even $J>2$ we do not have a reason to trust the eikonal model. Physically, this is because the terms generated in this regime are not Regge-enhanced, meaning that we do not have an argument as to why other contributions to the double discontinuity are subleading compared to the universal eikonal contribution. Using the integral expression above, it is easy to check that the eikonal ansatz generates the multi-stress tensor-like contributions
\be
\label{eq:reggetraj}
\tau_k(J) = (J-1)(d-2)+2k .
\ee
We do not expect such operators to be present in the full theory because multi-stress tensor-like operators develop anomalous dimensions. It would be interesting to understand the cancelation mechanism for these unwanted poles in the four-point function. This requires going beyond the simple eikonal model considered in this section.

\subsection{Stringy generation of the graviton pole}

As in flat space, we can also consider the stringy generation of the graviton pole. In this case, we consider the following Regge ansatz for the correlator \cite{Costa:2012cb,Costa:2017twz}
\be
{\rm dDisc} {\cal G}_{\text{str}} (z,\bar z) = -\pi^{d/2+1}\int_{-\infty}^\infty d\nu 4^{j(\nu)}\gamma(\nu, j(\nu)) \gamma(-\nu, j(\nu)) r\Big(\Delta(j(\nu)),j(\nu)\Big)j'(\nu)\sigma^{1-j(\nu)} {\Omega_\nu(\rho) \over 4 \nu} , 
\ee
where $r(\Delta,J)= \lambda_{\phi \phi{\cal O}_{\Delta,J}}\lambda_{\psi \psi {\cal O}_{\Delta,J}}K_{\Delta,J}$, where $K_{\Delta,J}$ can be found for example in \cite{Costa:2012cb}. As before, performing the $\sigma$-integral produces ${1 \over J-j(\nu)}$. Here, $j(\nu)$ is defined via the relation $\nu^2 + (\Delta(j(\nu))-d/2)^2=0$.

We can then close the $\nu$-contour to pick the pole at $J=j(\nu)$. In this way we get
\be
{I}(\rho,2) = 8 \gamma(-i d/2,2)\gamma(id/2,2) \lambda_{\phi \phi T_{\mu \nu}}\lambda_{\psi \psi T_{\mu \nu}}K_{d,2} {\cal H}_{d-1}(\rho) \ . 
\ee
One can check that this has precisely the same asymptotic as \eqref{eq:sumrhoI} at large $\rho$, and therefore it will correctly reproduce the graviton pole.
In the impact parameter space, the effect arises from ${\rm Im} \delta_{\text{str}}(S,L)$ as in flat space.

In fact, it is trivial to generalize the argument above to the whole leading Regge trajectory. We get in this case
\be
{I}(\rho,J) = {1 \over 2} 4^J \pi^{1+d/2} \gamma(i(\Delta-{d \over 2}),J)\gamma(-i(\Delta-{d \over 2}),J) \lambda_{\phi \phi {\cal O}_{\Delta,J}}\lambda_{\psi \psi {\cal O}_{\Delta,J} }K_{\Delta,J} {\cal H}_{\Delta-1}(\rho) \ . 
\ee
It is then trivial to check using \eqref{eq:generalint} that the expression above correctly generates the pole of the leading Regge trajectory operator with the correct residue
\be
c(\Delta,J) \simeq - {\lambda_{\phi \phi {\cal O}_{\Delta,J}}\lambda_{\psi \psi {\cal O}_{\Delta,J} } \over \Delta - \Delta_{{\cal O}_{\Delta,J}}} \ .
\ee
Again, this is identical to what happens in flat space and is in contrast to the eikonal scenario, which only correctly generates the graviton pole in the dispersion relations.

There is an interesting difference in characteristic graviton pole generation scales compared to flat space. Consider first the eikonal computation. At large impact parameters $L$ we have $\delta_\text{tree}(S,L) \sim G_N S e^{-(d-1)L}$, and therefore the graviton pole generation scale given by $\delta_\text{tree}(S_*,L) \sim O(1)$ corresponds to $S_* = {e^{(d-1)L} \over G_N}$ compared to a simple power in flat space. On the other hand, the stringy generation scale of the graviton pole corresponds to $S_* \sim e^{{L^2 \over \alpha'}}$ as before. Therefore, as in flat space, we expect that the eikonal generation of the graviton pole is relevant for $L \gtrsim L_\text{str} \log c_T$, where $L_\text{str} \sim {1 \over \Delta_\text{gap}}$, and $\Delta_\text{gap}$ is the gap in the spectrum of higher spin operators.

\subsection{Non-holographic theories}

In flat space, we derived a completely general sum rule, which the partial waves have to satisfy to correctly reproduce the graviton pole \eqref{eq:tauberianImpactb}. It would be interesting to derive an analogous sum rule in AdS. 

So far in this section, we have discussed mechanisms to generate the graviton pole using heavy operators with the relevant region in the inversion formula being $z, \bar z \to 0$. This mechanism is not unique.
To illustrate this, let us consider the correlator of $\phi_i \phi^i$ operators in the theory of free $N_s$  scalars in four dimensions. We have for the relevant part of the connected correlator \cite{Dolan:2000ut}
\be
{\cal G}_\text{conn}(z, \bar z) ={4 \over N_s} {z \bar z \over (1-z)(1-\bar z)} ,
\ee
where we normalized the two-point function to $\langle \phi_i \phi^i(x) \phi_i \phi^i(0) \rangle = {1 \over x^4}$. 

In this case, the double discontinuity is effectively localized at $\bar z=1$ in contrast to considerations earlier in the section. To apply the Lorentzian inversion formula, it is convenient to introduce the regulator ${z \bar z \over (1-z) (1-\bar z)} \to \Big({z \bar z \over (1-z) (1-\bar z)} \Big)^{1+\delta}$. We then get for the regulated double discontinuity of the correlator
\be
{\rm dDisc}{\cal G}_{\phi^2}^{\delta}(z, \bar z) = {8 (\sin \pi \delta)^2 \over N_s} \Big({z \bar z \over (1-z) (1-\bar z)} \Big)^{1+\delta} .
\ee
The inversion formula integral takes the form 
\be
c_\text{conn}(\Delta,J) &=\lim_{\delta\to 0}\frac{\Gamma \left(\frac{J+\Delta }{2}\right)^4}{4 \pi ^2 \Gamma (J+\Delta -1) \Gamma
   (J+\Delta )} \nn \\
   &\qquad\times \int_0^1 \int_0^1 \frac{dz}{z^2} \frac{d\bar z}{\bar z^2} \frac{(z-\bar z)^2}{(z\bar z)^2} G_{J+3,\Delta-3}(z,\bar z) {\rm dDisc}{\cal G}_{\phi^2}^{\delta}(z, \bar z),
\ee
where the relevant conformal block takes the following form
\be
G_{J+3,\Delta-3}(z,\bar z) &=  \frac{z\bar z}{z-\bar z}(k_{\Delta+J}(z) k_{4+J-\Delta}(\bar z)-k_{\Delta+J}(\bar z) k_{4+J-\Delta}(z)),\nn\\
k_\beta(z) &= z^{\beta/2} {}_2F_1 (\frac{\beta}{2},\frac{\beta}{2},\beta,z ).
\ee
The inversion integral now effectively factorizes and can be done using the following integral
\be
I_\a(h,p)&\equiv \int_0^1 \frac{dz}{z(1-z)} z^{\a} \left( \frac{z}{1-z} \right)^p k_{2h}(z)\nn\\
&= \frac{\G(\a+p+h)\G(-p)}{\G(\a+h)}{}_3F_2(h,h,\a+p+h;2h,\a+h;1).
\ee
The double zero in $(\sin \pi \delta)^2$  is canceled against the double pole in the following integral
\be
\lim_{\delta \to 0} (\sin \pi \delta)^2 I_{-1}(h,\delta) = \frac{\pi ^2 \Gamma (2 h)}{\Gamma (h)^2} \ . 
\ee
In this way, we get
\be
c_\text{conn}(\Delta,J) &=-\frac{2^{9-2 \Delta } (\Delta -2) (J+1) \Gamma \left(\frac{1}{2} (J-\Delta +5)\right) \Gamma \left(\frac{1}{2} (J+\Delta
   -2)\right)}{N_s (\Delta -J-2) (\Delta +J) \Gamma \left(\frac{1}{2} (J-\Delta +6)\right) \Gamma \left(\frac{1}{2}
   (J+\Delta -1)\right)}.
\ee
This expression has poles at the position of higher-spin operators with the correct residues given by the three-point function with higher spin currents \cite{Dolan:2000ut}
\be
c(\Delta,J) &\simeq - {\frac{8 \Gamma (J+1)^2}{N_s \Gamma (2 J+1)} \over \Delta-(2+J)}  . 
\ee
Therefore we see that a single Regge trajectory of conserved higher-spin currents dispersively generates itself in the dual channel. Moreover, the important contribution comes from \emph{the double light-cone region} $z,1-\bar z \to 0$ as opposed to \emph{the Regge region} $z, \bar z \to 0$ which was important for the eikonal and stringy models. This computation can be readily generalized to any other operators, free theories, and any number of dimension $d$. For the free scalar theory, we will get that the relevant part of the correlator takes the form $\Big( {z \bar z \over (1-z)(1-\bar z)} \Big)^{{d-2 \over 2}}$ and the double discontinuity is effectively localized at $\bar z=1$ in even number of dimensions, whereas in the odd number of dimensions all $0 \leq \bar z \leq 1$ contribute.

The graviton pole generation was also explored in the 3d Ising model \cite{Alday:2015ewa,Simmons-Duffin:2016wlq,Caron-Huot:2020ouj} and the $O(2)$ model \cite{Liu:2020tpf}. These cases are perhaps closer in spirit to the free field theory computation above, in the sense that including a few trajectories in the double discontinuity led to a reasonable prediction for the stress tensor pole (using non-rigorous extrapolation) and all $0\leq \bar z \leq 1$ contributed to the graviton pole in the Lorentzian inversion formula. In general, the integrand in the inversion formula is non-negative and therefore the problem is potentially amenable to the Tauberian analysis \cite{Pappadopulo:2012jk,Mukhametzhanov:2018zja,Pal:2022vqc}. It would be interesting to understand if a precise sum rule on the OPE coefficients necessary and sufficient to reproduce the graviton pole in a generic CFT can be derived along similar lines. 

\section{Discussion and future directions}\label{sec:discussions}

Twice-subtracted dispersion relations have long been recognized as a fundamental property of gapped relativistic quantum field theories. More recently, they have been successfully applied to gravitational theories, both in flat space and in AdS. An important new feature in this case is the presence of the graviton pole which on one hand controls gravitational attraction at long distances, and on the other hand is dispersive, namely it can be expressed via the dispersive integral of the discontinuity of the amplitude in flat space (or double discontinuity of the four-point correlator in AdS). In this paper, we explored different physical mechanisms for generating the graviton pole in dispersion relations. This question is particularly interesting because we expect that the graviton pole cannot be generated by QFT degrees of freedom \cite{Weinberg:1980kq,Caron-Huot:2024lbf}.

To address this question, we found it useful to consider experiments at different impact parameters $b$, and explore the characteristic energy scale $s_*(b)$ at which the graviton pole is generated. To define this scale, we used that partial waves must obey \eqref{eq:taubtheo} to generate the graviton pole. The crucial ingredient in the derivation of \eqref{eq:taubtheo} was going from a $1/t$ pole of the amplitude to the $1/\ea$ pole of the smeared amplitude. Then we took advantage of the positivity property of the functional \eqref{eq:defPsiIntro} for arbitrarily small $\ea>0$ to derive a sum rule that the nonnegative integrated discontinuity of the amplitude must satisfy.

In flat space, we identified two basic mechanisms to generate the graviton pole. At large impact parameters, which is responsible for the $t \to 0$ limit of the amplitude, we found that the eikonal scattering correctly reproduces the graviton pole. This mechanism is completely universal, and we expect it to be realized in any theory of gravity. At smaller impact parameters related to the string length $\ell_s$, however, the tree-level gravitational scattering amplitude can be generated by weakly coupled higher-spin resonances, see \figref{fig:ACV}. This second mechanism is particular to string theory and is related to the fact that, in this case, there exists a hierarchy of scales $M_{Pl} \gg M_{s}$. We have observed that the gravitational EFT breaks down below the impact parameter $b_\star$ for which the effective classical spacetime dimension ${\cal D}(b) = {\partial \log s_*(b) \over \partial \log b}$ departs from its approximately constant semi-classical value, see \figref{fig:gravScaleSch}.

Going to the case of $AdS_{d+1}/CFT_d$, requires minimal modifications related to different dependence of various formulas on the impact parameters when it becomes larger than the AdS curvature $R_{AdS}$. We explicitly checked the AdS eikonal and stringy mechanisms to generate the graviton pole.  We also discussed non-holographic theories, such as the free CFTs, in which case the graviton pole generation is not given by a simple physical picture of high-energy scattering in AdS.  It would be interesting to derive a universal Tauberian theorem for the graviton pole, or, equivalently, the stress tensor exchange in the OPE, in AdS/CFT. 

We explored some consequences of the graviton pole sum rules in the context of high-energy gravitational scattering. We have seen in \secref{sec:blackdiscBH} that black hole production leads to the one-loop non-analytic term in the amplitude.  In some theories, it is related by crossing and unitarity to the graviton pole and therefore is completely fixed. Assuming that black holes are the only source of this correction leads to an estimate of the size of the collapse region in agreement with the classical estimate. Similarly, we have used the twice-subtracted dispersion relations to bound from above the amount of scattering in the string/black hole transition region \secref{sec:greydisc}.
It would be very interesting to understand if this simple analysis can be refined further.

There are several obvious questions that we have not addressed in the paper. First, we did not cover perhaps the most interesting cases: four-dimensional asymptotically flat and de Sitter spacetimes. In four-dimensional asymptotically flat spacetime, the infrared structure of the gravitational theory requires re-formulation of scattering theory \cite{Strominger:2017zoo,Hannesdottir:2019opa,Prabhu:2022zcr,Prabhu:2024lmg}, and at the moment it is not known what substitutes the standard dispersion relations in this case. A convenient nonperturbative tool to study this case is to consider dispersion relations in $AdS_4$ and take the flat space limit \cite{Caron-Huot:2021rmr}. 
The AdS radius  $R_{AdS}$ enters various bounds and plays the role of the IR regulator. It is, however, not yet clear how to systematically apply this approach, for example, to our universe. Related to the last comment, we observed that in de Sitter there exists a large impact parameter obstruction to dispersion relations due to the flip of the sign of the Shapiro time delay at the cosmological scale $R_{dS}$, see \eqref{eq:phaseshiftdispersion} and the discussion around it. Second, it would be very interesting to generalize our analysis beyond the tree-level, or, equivalently, $O(G_N)$. We know that the IR amplitude admits expansion in powers of $G_N$,\footnote{These powers do not always have to be an integer, see e.g. \cite{Binder:2019mpb,Chester:2023qwo}.} moreover, some of the terms that appear in the loops are completely fixed by unitarity and crossing. We have seen that already at one loop some terms of this type receive contributions from the black hole production region.  Are there further lessons to be learned from the structure of the perturbative loop corrections about UV physics?\footnote{See, for example, \cite{Mukhametzhanov:2018zja} where Tauberian theorems were used to capture subleading terms in the context of CFT and meromorphic amplitudes. The difference here is that corrections in $\ea$ are analytic, and thus the methods of \cite{Mukhametzhanov:2018zja} are not directly applicable.} Generalizing the analysis of \cite{Guerrieri:2021ivu,Guerrieri:2022sod} to the regime $\ell_s \gg \ell_{Pl}$ would be an important step in this direction, see e.g. \cite{Acanfora:2023axz,Chen:2022nym,EliasMiro:2022xaa} for for analogous results in the context of QFT without gravity. Relatedly, it would be interesting to see if one can improve the existing bounds on the IR observables by subtracting the known part of the dispersive integral that generates the graviton pole, e.g. the eikonal part. The ability to do so seems to be closely related to the notion of infrared causality, see e.g. \cite{deRham:2022hpx}. 

Another question concerns possible phases of gravity beyond the ACV picture of scattering in weakly coupled string theory displayed in \figref{fig:ACV}. Or, relatedly, are there other physical mechanisms to generate the graviton pole in the twice-subtracted dispersion relations beyond what we considered in the paper? From the recent bootstrap analysis, we can bring up at least two nontrivial examples of this type. One concerns the tree-level string-like scattering. It was found in \cite{Albert:2024yap} that there exist amplitudes that unitarize the graviton in the same way as tree-level string theory does but contain only a single clearly detectable trajectory.\footnote{It could be that the subleading trajectories are simply very small in agreement with the analysis of \cite{Eckner:2024pqt}.} It is not known what is the physical setting in which amplitudes of this type arise. Another example is the unitarization of the graviton pole thanks to a single trajectory of Planckian resonances dubbed graviballs in \cite{Blas:2020och,Blas:2020dyg,Guerrieri:2021ivu,Guerrieri:2022sod}. This mechanism could be possibly realized in string theory when $M_s \sim M_{Pl}$. Finally, we have seen that in free CFTs the graviton pole is generated by a single Regge trajectory of higher spin conserved currents.

\section*{Acknowledgments}

We would like to thank Jose Calderon-Infante, Andrea Guerrieri, Giulia Isabella, Robin Karlsson, Leonardo Rastelli,  Riccardo Rattazzi, Anna Tokareva, Piotr Tourkine, Irene Valenzuela, Gabriele Veneziano, and Zahra Zahraee for useful discussions. We also would like to thank the participants of the Bootstrap 2024 meeting in Madrid, where some of the material contained in the paper was presented,\footnote{\url{https://teorica.fis.ucm.es/bootstrap2024/programW2.html}} for valuable comments. This project has received funding from the European Research Council (ERC) under the European Union’s Horizon 2020 research and innovation program (grant agreement number 949077). The work of KH is supported by the Simons Foundation grant 488649 (Simons Collaboration on the Nonperturbative Bootstrap), by the Swiss National Science Foundation through the project 200020 197160, through the National Centre of Competence in Research SwissMAP and by the Simons Collaboration on Celestial Holography.

\appendix
\section{Kinematics, unitarity, partial waves and all that}\label{sec:convention}
In this appendix, we briefly review the definitions and properties of the amplitude used in this work. We used the same convention as in \cite{Correia:2020xtr} and refer the reader to this reference for extensive details.

\paragraph{Kinematics}\mbox{}\\
In this work, we focus on the two-to-two scattering of nonidentical massive scalars $A,B\to A,B$ with the same mass $m$. The scattering amplitude is given by a function of two variables $T(s,t)\equiv T^{A,B\to A,B}(s,t)$ where $s,t$ are the usual Mandelstam variables $s+t+u=4m^2$. $s$ is the square of the center of mass energy and $t$ is the square of the momentum transfer. They are related to the scattering angle via the usual relation
\begin{equation}
    z\equiv \cos\theta = 1+ \frac{2t}{s-4m^2}\,.
\end{equation}

\paragraph{Partial wave}\mbox{}\\
The amplitude can be decomposed into a complete set of partial waves 
\begin{equation}\label{eq:PWExpansion}
    T(s,t) = \frac{1}{2}\sum_{J=0}^{\infty}\njd f_J(s) \Pjd\left(1+ \frac{2t}{s-4m^2}\right)\,,
\end{equation}
where $f_J(s)$ are the partial waves and describe the dynamics of the scattering process; $\njd$ is a normalization coefficient that we will define shortly and we will define $\njdnID=\njd/2$  to not clatter the equations. $\Pjd(z=\cos\theta)$ are the Legendre polynomials in $d$ dimensions
\begin{equation}
    \Pjd(z)= {}_2F_1\left(-J,J+d-3, \frac{d-2}{2}; \frac{1-z}{2}\right) = \frac{C_J^{\left( \frac{d-3}{2} \right)}(z)}{C_J^{\left( \frac{d-3}{2} \right)}(1)}\,,
\end{equation}
where $C_J^{\left( \alpha \right)}(z)$ are the usual Gegenbauer polynomials. They satisfy the orthogonality relation
\begin{equation}
    \frac{1}{2}\int\limits_{-1}^1 d z\, (1-z^2)^{{d-4 \over 2}} \Pjd(z) P^{(d)}_{\tilde J}(z) =\frac{\delta_{J \tilde J}}{ {\cal N}_d\, n_J^{(d)}}
\end{equation}
and the completeness relation
\begin{equation}
\label{eq:PP}
\sum_{J=0}^\infty n_J^{(d)} P^{(d)}_J(y) P^{(d)}_J(z) = {2 \over \mathcal{N}_d} (1 - z^2)^{4-d \over 2} \delta(y - z)\, ,
\end{equation}
with
\begin{align}\label{eq:nJd2}
\cN_d&=\frac{(16\pi)^{\frac{2-d}{2}}}{\Gamma (\frac{d-2}{2} )}\, , ~~~n_J^{(d)}=\frac{(4\pi)^{\frac{d}{2}}(d+2J-3) \Gamma (d+J-3)}{\pi\,\Gamma \left({\frac{d-2}{2}}\right) \Gamma (J+1)}\ .
\end{align}
The partial wave expansion \eqref{eq:PWExpansion} can be inverted
\begin{equation}
    f_J(s) = \cN_d\int_{-1}^{1}dz (1-z^2)^{\frac{d-4}{2}}\Pjd(z)T(s,t(z))\,.
\end{equation}
For a gravitational amplitude, the partial wave expansion converges pointwise for $-1< z< 1$ in $d>5$ and as a distribution in $d=5$, see \cite{Haring:2022cyf}.

For the smeared amplitude \eqref{eq:SmearedDef}, with a general smooth functional $\psi_{\ea,\eb}$ whose behavior close to the end points is
\begin{equation}\label{eq:psiabGen}
    \begin{split}
         \psi_{\ea,\eb}(q)&\overset{q\to 0}{\sim} q^\ea\,,~~~ \ea>0\\
    \psi_{\ea,\eb}(q)&\overset{q\to q_0}{\sim} (q_0-q)^\eb\,,~~~ \eb\geq 0\,,
    \end{split}
\end{equation}
the decomposition in partial wave converges in $d\geq 5$ and reads
\begin{equation}\label{eq:smearedPWdef}
    \Tpsiab(s) = \frac{1}{2}\sum_{J=0}^{\infty}\njd f_J(s)\Pjd[\psi_{\ea,\eb}]\,,
\end{equation}
where we defined 
\begin{equation}
    \Pjd[\psi_{\ea,\eb}] \equiv \int_{0}^{q_0} dq\,q \left[\psi_{\ea,\eb}(q)\right] \Pjd\left(1+ \frac{2t}{s-4m^2}\right)\,.
\end{equation}
In writing \eqref{eq:smearedPWdef}, we swapped the sum and integral, this step was justified in \cite{Haring:2022cyf}. In the main text, we will mostly use $\eb=\frac{d-1}{2}$.

\paragraph{Unitarity}\mbox{}\\
Nonperturbative unitarity is conveniently expressed at the level of the partial waves $f_J(s)$. It is instructive to define the following combination \cite{Paulos:2017fhb}
\begin{equation}
    S_J(s)= 1+ i \frac{(s-4m^2)^{\frac{d-3}{2}}}{\sqrt{s}} f_J(s) = 1+ i a_J(s)\,.
\end{equation}
Then unitarity states 
\begin{equation}\label{eq:unitarity}
    |S_J(s)|\leq 1\,,~~\text{ or }~~~ 2 \Im f_J(s)\geq \frac{(s-4m^2)^{\frac{d-3}{2}}}{\sqrt{s}} |f_J(s)|^2\,,~~ s\geq 4m^2 
\end{equation}
This can be solved in therm of the phase shift $\delta_J(s)$
\begin{equation}
    S_J(s) = e^{2i \delta_J(s)} \,,~~ \text{with }~~  \Im\delta_J(s)\geq 0\,.
\end{equation}

\paragraph{Impact parameter representation}\mbox{}\\
When considering the large-energy and large-spin regime of the scattering process, it is instructive to consider the impact parameter representation of the amplitude, see e.g. \cite{DiVecchia:2023frv,Bellazzini:2022wzv}. To this end, we consider the double scaling limit 
\begin{equation}
    J\to\infty\,,~s\to\infty\,,~~ b=\frac{2J}{\sqrt{s-4m^2}} ~\text{fixed}\,.
\end{equation}
Where by using the usual relation $\vec J=\vec b\times\vec p$, $b$ has the interpretation of distance in the transverse direction. In this limit, starting from the partial wave expansion \eqref{eq:PWExpansion}, we get the impact parameter representation  
\begin{align}
    T(s,t=-q^2)&\simeq 2 i s(2\pi)^\frac{d-2}{2}\int_{0}^{\infty} db\, d^{d-3} (bq)^{\frac{4-d}{2}} J_{\frac{d-4}{2}}(bq)\left(1-e^{2i\delta(s,b)}\right) \label{eq:impactParDef}\\
    &\simeq 2 i s\int d^{d-2}\vec b\, e^{-i\vec q\vec b}\left(1-e^{2i\delta(s,|\vec b|)}\right)\,.
\end{align}
In the second line, we introduced the transverse vector $\vec b$ to highlight that it is nothing else than the Fourier transform of the amplitude in the transverse direction. We also defined the phase shift in impact parameter space by 
\begin{equation}
    \delta_{J=\frac{b\sqrt{s-4m^2}}{2}}(s)= \delta(s,b) +\dots \,, ~~~ s\to \infty\,,
\end{equation}
where $\dots$ are suppressed at large $s$.

\section{Unitarity bound on the nonperturbative Regge behavior}\label{sec:nonpretUnitRegge}
In this section, we derive a simple bound on the nonperturbative Regge limit of the scattering amplitude assuming the amplitude is controlled by an isolated singularity in the complex spin. The result derived in this section is agnostic whether the theory is gapless, gapped, or gravitational.

\vspace{5pt}
{\bf Theorem:} Let us assume that the leading Regge behavior of the elastic nonperturbative $2\to 2$ scattering amplitude in the interval $- t_0 \leq t \leq 0$ comes from a singularity at $j(t)$ (for example, a Regge pole, a pair of complex conjugate Regge poles, or more complicated non-analyticity). Then 
\begin{equation}\label{eq:j0Def}
    j_0=\sup_{- t_0 \leq t \leq 0} \Re j(t) \leq 1\,.
\end{equation}

Consider first the case of a Regge pole where the scattering has the following asymptotic
\begin{equation}\label{eq:T_ReggePole}
    \lim_{|s| \to \infty}T(s,t) \simeq f(t) (-i\alpha' s)^{j(t)} , ~~~ - t_0 \leq t \leq 0 .
\end{equation}
The proof is based on the partial wave expansion for the smeared amplitude \eqref{eq:smearedPWdef}
that we recall here
\begin{equation}
    \Tpsiab(s) = \sum_{J=0}^{\infty}\njdnID f_J(s)\Pjd[\psi_{\ea,\eb}]\,.
\end{equation}
Here we consider a general functional whose behavior close to the end point is given by \eqref{eq:psiabGen}.
Unitarity, see \eqref{eq:unitarity}, implies that 
\begin{equation}
    |f_J(s)|\leq \frac{2\sqrt{s}}{(s-4m^2)^{d-3\over 2}}\lesssim s^{4-d\over 2}\,.
\end{equation}
Then using the fact that
\begin{equation}\label{eq:PjpsiBound}
    \Pjd[\psi_{\ea,\eb}]\leq \min\left(C_1,C_2 \max\left(\left({J \over \sqrt{s}}\right)^{-2-\ea},\left( {J \over \sqrt{s}}\right)^{\frac{1-d}{2}-\eb} \right) \right) \,,
\end{equation}
we get, by splitting the sum in the partial wave expansion 
\begin{align}
    |\Tpsiab(s)|&\lesssim s^{4-d\over 2}\left(C_1 \sum_{J=0}^{b_0\sqrt{s}} \njd + C_2 \sum_{J=b_0\sqrt{s}+1}^{\infty}\njd \max\left(\left({J \over \sqrt{s}}\right)^{-2-\ea},\left( {J \over \sqrt{s}}\right)^{\frac{1-d}{2}-\eb} \right)\right)\\
    &\lesssim C_{\ea,\eb}^{A,B\to A,B} s \,, ~~ \text{for } \ea\geq d-4\,,\, \eb \geq \frac{d-3}{2}\,.
\end{align} 
The last line is obtained only by assuming unitarity and restricting to a class of smearing functions with $\ea \geq d-4$ and $\eb \geq \frac{d-3}{2}$. This argument can be trivially extended to the integral in the interval $\tilde{q}_0 \leq q \leq q_0$
\begin{equation}
    \int_{\tilde q_0}^{q_0} dq q \psi_{\ea,\eb}(q) T(s, t=-q^2) =  \int_{0}^{q_0} dq q \psi_{\ea,\eb}(q) T(s, t=-q^2) -  \int_0^{\tilde q_0} dq q \psi_{\ea,\eb}(q) T(s, t=-q^2)\,.
\end{equation}
Choosing a functional such that $\psi_{\ea,\eb}(q) \sim (\tilde{q}_0 - q)^\eb$, this integral is also bounded by $s$.

Consider now applying the smearing to the asymptotic behavior of a single Regge pole \eqref{eq:T_ReggePole}
\begin{equation}
    \int_{\tilde q_0}^{q_0} dq q \psi_{\ea,\eb}(q)   f(-q^2) (-i\alpha' s)^{j(-q^2)} \sim s^{j_0}\,
\end{equation}
where $j_0$ is defined in \eqref{eq:j0Def} and $\sim$ here means asymptotic up to a slow-growing function such as $\log s$. These functions can arise from integrating $q$ around $j(-q^2)=1$. Using the fact that nonperturbative unitarity bounds the LHS by $s$, it follows that 
\begin{equation}\label{eq:j0Bound}
    j_0\leq 1\,.
\end{equation}
The argument is not changed if we model the asymptotic behavior by a complex pair of Regge poles or more complicated isolated non-analyticity in the $j$-plane. In the latter case, the function $f(t)$ is replaced by $g(s,t)$, where $g(s,t)$ is a function of slow variation in $s$.\footnote{Recall that a function of slow variation is defined by $\lim\limits_{s\to \infty}\frac{g(\lambda s,t)}{g(s,t)}=1$ \cite{Qiao:2017xif}. The logarithm is an example of such a function. This is what happens if the amplitude is dominated by a Regge cut. In such a case, the bound is satisfied up to a slow-growing function.}

Let us make a couple of remarks about the bound derived above. First, the bound \eqref{eq:j0Bound} \emph{does not} mean that 
\begin{equation}\label{eq:sboundT}
    \lim_{|s|\to\infty} \frac{|T(s,t)|}{|s|}\lesssim \text{cst.}\,,
\end{equation}
Rather, it means that a simple singularity in the $j$-plane with intercept $j_0>1$ is not compatible with nonperturbative unitarity. Indeed, the eikonal amplitude satisfies unitarity, evades \eqref{eq:sboundT}, and behaves as \eqref{eq:Reggelimiteikonal}. 

Interestingly, for gapped theories, a scattering amplitude saturating the Froissart-Martin bound $T(s,0) \lesssim s (\log s)^{d-2}$ is precisely of the type \eqref{eq:T_ReggePole}, where $f(t)$ is replaced by a function of slow variation in $s$. Similarly, at fixed momentum transfer we have the following bound \cite{Eden:1971fm}
\begin{equation}
    T(s,t<0)\lesssim s (\log s)^{\frac{d-1}{2}}\,,
\end{equation}
and therefore the bound \eqref{eq:j0Bound} is again saturated. This can be obtained, for example, by an amplitude dominated by a Regge cut. See e.g. \cite{Kupsch:1972nf,Kupsch:1982aa,Kupsch:2008hq} for interesting work towards constructing an amplitude saturating the Froissart bound.

Let us remark that the bound derived here relies on the bound on the smeared Legendre polynomials \eqref{eq:PjpsiBound}. This is not a proven property but has been heavily checked numerically. A sharp proof of this statement would place this derivation on a stronger footing.

{\bf Corollary:} In $d>4$ quantum gravity the nonperturbative Regge limit is never controlled by an isolated singularity in the $j$-plane.

\section{The space of positive functionals}\label{app:posFunctional}
The space of functionals that one can consider is in general infinite and nontrivial. Here we will focus on the polynomial functionals of the form 
\begin{equation}\label{eq:psiAppChar}
    \psi_{\ea,\eb}(q) = q^\ea (q_0-q)^{\eb}\,.
\end{equation}
To start, a necessary condition to obtain positivity is to look at \eqref{eq:PositivityCpsi} in the large $x', J$ limit, i.e. in the impact parameter space. Then, let us first consider the large $x,y$ region where \eqref{eq:conditionPosCpsi} becomes
\begin{equation}
    \widehat{\psi}(b) = \int_0^{q_0}dq\ q\psi(q)\ (qb)^{\frac{4-d}{2}}J_{\frac{d-4}{2}}(bq) \geq 0~~ \forall~b \,,
\end{equation}
which is the Fourier transform of $\psi(q)/q^{d-4}$.  Below, we will set $q_0=1$ to not clatter the equations.

The integral can be performed exactly, and at large $b$ we obtain
\begin{equation}
    \begin{split}
    \int_0^{1}dq\ q q^\ea(1-q)^\eb&\ (qb)^{\frac{4-d}{2}}J_{\frac{d-4}{2}}(bq)\\
   &= b^{-\ea-2}\frac{ 2^{\ea-\frac{d}{2}+3} \Gamma \left(\frac{\ea}{2}+1\right)}{\Gamma \left(\frac{1}{2}
   (d-\ea-4)\right)}-b^{-\ea-3}\frac{\eb2^{\ea-\frac{d}{2}+4} \Gamma \left(\frac{\ea+3}{2}\right)}{\Gamma \left(\frac{1}{2}
   (d-\ea-5)\right)}\\
   &+b^{-\eb-\frac{d-1}{2}} \sqrt{\frac{2}{\pi }} \Gamma (\eb+1)  \cos
   \left(b-\frac{1}{4} \pi  (2 \eb+d-1)\right)+\dots    
    \end{split}
\end{equation}
The third term oscillates at large impact parameter space. Thus to obtain positivity, a necessary condition is $\ea< \frac{1}{2}(d-5+2\eb)$. It is also necessary that the coefficient of the leading term is positive which implies that $\ea< d-4$.  As $\ea\to d-4$, the first term vanishes and positivity could be broken at finite impact parameters. Thus, we require that the second term dominates the third; in other words, we require $\eb\geq \frac{d-1}{2}$. Finally, it is possible to check that positivity remains satisfied for $\ea= d-4$ if $\eb\geq \frac{d-1}{2}$. All in all, we obtain the condition,
\begin{equation}
    \ea\leq d-4\,, ~~ \eb \geq \frac{d-1}{2}\,.
\end{equation}
In the main text, we always chose $\eb$ to saturate this inequality and use $\ea$ as a parameter. Note that the discussion above would not change if we consider $\psi_{\ea,\eb}\to q^{\ea}(q_0-q)^\eb Q(q)$, if $Q(q)$ does not have zeros at $q=0,q_0$.  

The condition so far is necessary to have a positive functional that contains the large $x,y$ region. The exact region where \eqref{eq:PositivityCpsi} is satisfied depends on the dimension and the choice of $\ea\,,\eb$ and $Q(q)$. We have observed that for the class of functionals \eqref{eq:psiAppChar}, in $d\geq 6$ the limiting case is given by $J=0$ which leads to a large region in $x,y$ where positivity is satisfied, see \figref{fig:RegionPositive}. In $d=5$, more spins have to be considered but the qualitative picture remains. 

\section{Tauberian theorems}\label{app:tauberian}
In this appendix, we review Tauberian theorems used in \secref{sec:TauberianTHMDerivation} to derive a sum rule for the graviton pole. Our goal is not to provide a comprehensive review of the subject but rather to equip the reader with the necessary tools to follow the derivation. For more detailed information, we refer the reader to \cite{korevaar_tauberian_2004}. In the main text, we employed the Hardy-Littlewood theorem for the Laplace transform and mentioned the Wiener-Ikehara Tauberian theorem, which we state below without proof.

\paragraph{Hardy-Littlewood theorem for Laplace transform}\mbox{}\\
Let $f(t)$ be a non-negative distribution and $F(s)$ its Laplace transform 
\begin{equation}
F(s) = \int_0^\infty d\omega \, f(\omega) e^{-s\omega} \,,
\end{equation}
exists for $s>0$. Suppose that for some constant $\a\geq 0$
\begin{equation}
	F(s)\simeq \frac{C}{s^\alpha} \, , \, s\to 0 \, ,
\end{equation}
where $A(s)\simeq B(s), ~ s\to 0$ means $\lim_{s \to 0}{A(s) \over B(s)} = 1$.
Then 
\begin{equation}
	\int_0^T d\omega \, f(\omega) \simeq \frac{C}{\G(\alpha+1)} T^\alpha \, , ~~ T \to \infty \,.
\end{equation}
This theorem was used to constrain the growth of the average impact parameter amplitude in \eqref{eq:tauberianImpactb}.

\paragraph{Wiener-Ikehara Tauberian theorem for Dirichlet series}\mbox{}\\
Let 
\begin{equation}
F(s) = \sum_{n=1}^\infty a_n n^{-s} \ ,
\end{equation}
be a Dirichlet series with $a_n\geq 0$. If $F(s)$ is analytic in $\Re(s)> s^*$ for some $s^*\in \mathbb{R}$ and behave as a simple pole at $s=s^*$
\begin{equation}
	F(s) \simeq \frac{C}{s-s^*} \, , \, ~~ s\to s^* \, ,
\end{equation}
then
\begin{equation}
	\sum_{n=1}^{N} a_n \simeq \frac{C}{s^*} N^{s^*}  \,, ~~N \to \infty  \,.
\end{equation}
This theorem could `almost' be used in the fixed spin discussion at the end of \secref{sec:taubJdiscussion}.

\section{Eikonal scattering and partial wave expansion}\label{sec:AppetizerEikonalModel}

We want to check partial wave unitarity for the eikonal model \eqref{eq:treeleveldisc}, where we set $m=0$. Curiously, the projection integral can be performed exactly,  \cite{Muzinich:1987in}, and the result takes the following form\footnote{We thank Giulia Isabella and Piotr Tourkine for discussions on related topics.}
\be
S_J(s) = \sqrt{s} \int_0^\infty db J_{d-3+2J}( b\sqrt{s}) e^{2 i \delta(s,b)} . 
\ee
This formula admits an interesting large $J \to \infty$ limit for which the integral effectively localizes at $J = {b \sqrt{s} \over 2}$ and we get
\be
\label{eq:largeJeikmodel}
S_J(s) \simeq e^{2 i \delta(s,b)}, ~~~ J = {b \sqrt{s} \over 2} \gg 1 . 
\ee
We then get that the unitarity condition \eqref{eq:unitarity} for the phase shift becomes
\be
{\rm Im} \delta(s,b) \geq 0 \ . 
\ee

\begin{figure}[htbp]
    \centering
    \includegraphics[width=0.6\textwidth]{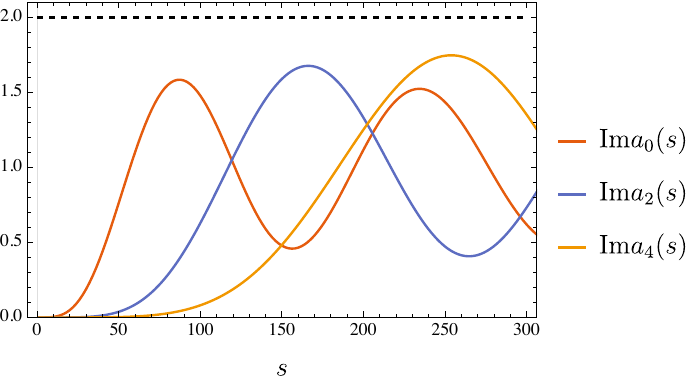}
    \caption{The imaginary parts of partial waves ${\rm Im} a_J(s)$ in the eikonal model in $d=7$. We set $8 \pi G_N = 1$. The black dashed line curve $2$ presents a nonperturbative unitarity bound. Higher spin partial waves come closer and closer to saturating it in agreement with \eqref{eq:largeJeikmodel}. The maximal value of the spin $J$ partial wave is reached at energies $\sim J^{2 {d-4 \over d-2}}$.}
    \label{fig:PWToy}
\end{figure}

Let us explore partial wave unitarity for low $J$ next. Notice that $|S_J| \leq 1$ in particular implies that $2 \leq {\rm Im} a_J(s) \leq 0$ 
\be
{\rm Im} a_J(s) = 2 \sqrt{s} \int_0^\infty J_{d-3+2J}( b\sqrt{s}) \sin^2 \delta_\text{tree}(s,b) d b . 
\ee
These partial wave projections can be computed explicitly. We plot the first few in \figref{fig:PWToy}. We find that they satisfy nonperturbative unitarity.

\newpage
\bibliographystyle{JHEP}
\bibliography{refs}

\end{document}